\documentclass[
 reprint, superscriptaddress,
 nofootinbib,
 amsmath,amssymb,
 aps,
prx,
]{revtex4-2}

\usepackage{graphicx}
\usepackage{dcolumn}
\usepackage{bm}
\usepackage{hyperref}
\usepackage{braket}
\usepackage{upgreek}
\usepackage{enumitem}
\usepackage{xcolor}
\usepackage{stmaryrd}
\usepackage{placeins}

\definecolor{mygreen}{HTML}{59B51C}

\begin{document}

\title{Detuning- and Stark-robust Rydberg gates}
\author{Elie Bataille$^{\,\dagger}$}
\thanks{These authors contributed equally to this work.}
\affiliation{California Institute of Technology, Pasadena, CA 91125, USA}
\author{Gyohei Nomura}
\thanks{These authors contributed equally to this work.}
\affiliation{California Institute of Technology, Pasadena, CA 91125, USA}
\author{Manuel Endres$^{\,\ddagger}$}
\affiliation{California Institute of Technology, Pasadena, CA 91125, USA}
\affiliation{Oratomic, Pasadena, CA 91125, USA}

\begin{abstract}
Rydberg entangling gates driven by a two-photon transition in alkali atoms suffer from an adverse scaling of light-shift-induced errors. Robustness to such detuning errors is known to be impossible to achieve in the design of conventional Rydberg gate protocols, where only one of the qubit states is coupled to the Rydberg state. Here, we show that in a more general framework, in which both qubit states take part in the gate, full or partial robustness to these errors can be realized. We present two gate constructions, which either cancel the errors outright or convert them into single-qubit errors that can be corrected locally. We map the regimes -- in terms of light-shift strength, intensity inhomogeneity, and Rydberg decay rate -- in which these protocols outperform the widely used time-optimal Rydberg gate, and find that they already include the conditions of state-of-the-art experiments. Finally, we show the existence of a Rydberg `fly-by' entangling gate, an important primitive for an emerging class of neutral-atom quantum computing architectures.
\end{abstract}

\maketitle
\begingroup
\renewcommand{\thefootnote}{\ensuremath{\dagger}}\footnotetext{bataille@caltech.edu}%
\renewcommand{\thefootnote}{\ensuremath{\ddagger}}\footnotetext{mendres@caltech.edu}%
\endgroup

\section{Introduction}

Neutral atom systems are one of the leading modalities for building quantum computers~\cite{bluvsteinQuantumProcessorBased2022,grahamMultiqubitEntanglementAlgorithms2022,bluvsteinLogicalQuantumProcessor2024,Reichardt2024FaulttolerantNeutralatom,Rines2025LogicalArchitectureMotion,Bluvstein2026FaulttolerantNeutralatomArchitecture}, due to their growing scale~\cite{gygerContinuousOperationLargescale2024,Norcia2024IterativeAssemblyYb,Chiu2025ContinuousOperationCoherent,Manetsch2025TweezerArray6100,Wang2026Trapping11000Atoms}, decreasing error rates~\cite{everedHighfidelityParallelEntangling2023,finkelsteinUniversalQuantumOperations2024,Radnaev2025UniversalNeutralatomQuantum,tsaiBenchmarkingFidelityResponse2025,Muniz2025HighfidelityUniversalGates,evered2026highfidelityentanglinggatesnonlocal}, and transport-enabled connectivity that is capable of highly efficient fault-tolerant computation~\cite{xuConstantoverheadFaulttolerantQuantum2024,Cain2026ShorsAlgorithmPossible}. Besides encoded computation, digital neutral atom computers have enabled advances in digital quantum simulation~\cite{bluvsteinQuantumProcessorBased2022,Evered2025ProbingKitaevHoneycomb} and quantum-enhanced metrology~\cite{finkelsteinUniversalQuantumOperations2024,Cao2024MultiqubitGatesSchrodingerCat,shawMultiensembleMetrologyProgramming2024}.
Across these endeavors, entanglement is generated by transiently exciting one of the qubit states to a Rydberg state, in so-called \emph{Rydberg gates}~\cite{jakschFastQuantumGates2000,saffmanQuantumInformationRydberg2010,saffmanQuantumComputingAtomic2016}.

The fidelity of Rydberg gates has improved steadily thanks to efficient protocols~\cite{levineParallelImplementationHighFidelity2019,janduraTimeOptimalTwoThreeQubit2022,paganoTO}, a detailed understanding of error sources~\cite{deleseleucAnalysisImperfectionsCoherent2018,Jiang2023SensitivityQuantumGate,everedHighfidelityParallelEntangling2023,Peper2025SpectroscopyModelingYbRydberg,tsaiBenchmarkingFidelityResponse2025}, and practical progress in laser phase-noise suppression~\cite{levineHighFidelityControlEntanglement2018,Li2022ActiveCancellationServoInducedNoise,Denecker2025MeasurementFeedforwardCorrection}, laser power, and beam shaping~\cite{schroffAccurateHolographicLight2023}. Because Rydberg gates are the main contributor to the error rate of quantum error correction cycles in leading experiments~\cite{Bluvstein2026FaulttolerantNeutralatomArchitecture}, reducing their error further is important for progress in all of the above applications. Since transporting atoms in and out of Rydberg zones is costly, scaling these gates with high fidelity -- to larger zones through holographic beam shaping or through time multiplexing -- is a related and similarly important challenge.

In alkali atoms, typical two-photon excitation schemes produce very large light shifts through the AC Stark effect, which require a laser-intensity homogeneity and stability that could be challenging for holographic beam shaping or beam rastering. As an example, the current state-of-the-art fidelity for a neutral-atom entangling gate, 99.85\%, was obtained in a single one-dimensional array of 8 gate sites with Gaussian-shaped Rydberg beams~\cite{evered2026highfidelityentanglinggatesnonlocal}, while the highest gate fidelity reported in a more general two-dimensional alkali-atom array with a holographically generated flat-top beam stands at 99.5\%~\cite{Bluvstein2026FaulttolerantNeutralatomArchitecture}. Crucially, light-shift-induced detuning errors are the asymptotically dominant error source as laser power increases: they constrain state-of-the-art experiments to operate away from their otherwise optimal parameters, curtailing the achievable fidelity. As a consequence, at a 1\% root-mean-square (RMS) intensity inhomogeneity -- a level matching the state of the art of typical holographic beam-shaping setups -- reaching entangling-gate fidelities close to 99.9\% with two-photon Rydberg schemes in alkali atoms might be challenging, regardless of the available laser power. Other practical challenges to scaling gate fidelity across larger processors, such as electric-field gradients, would likewise benefit from protocols with a reduced sensitivity to detuning errors.

Numerous works have studied gate protocols that reduce the fidelity sensitivity to other error sources, such as pure amplitude errors~\cite{beterovRydbergBlockadeForster2015,janduraOptimizingRydbergGates2023,fromonteilProtocolsRydbergEntangling2023,Mohan2023RobustControlOptimalRydberg,Xue2024HighFidelityRobustControlledZ,Li2025ErasingDopplerDephasing,kozenko2026numericallyoptimizedamplituderobust}. However, Ref.~\cite{janduraOptimizingRydbergGates2023} has shown that no \emph{conventional} Rydberg gate protocol -- defined here as a protocol where only $\ket{1}$ is coupled to the Rydberg state $\ket{r}$ while $\ket{0}$ remains a spectator -- can suppress the sensitivity to detuning errors.

\begin{figure}
\centering
    \includegraphics[width=86mm]{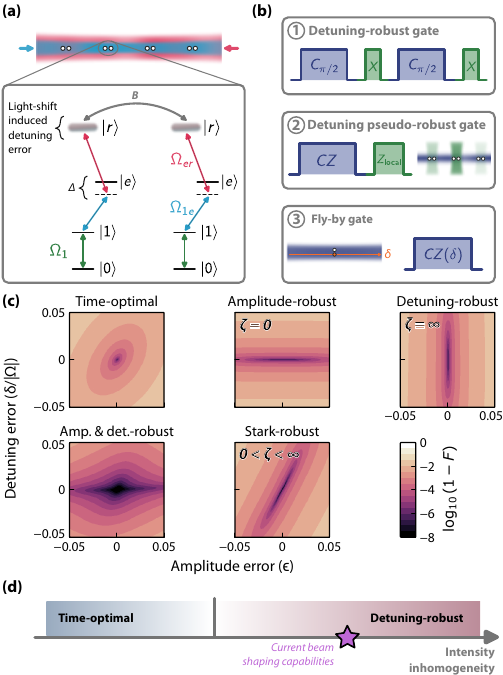}
    \caption{\label{fig:overview}\textbf{Overview of the physical setup and results.} (a) Two-atom level structure for two-photon Rydberg excitation, with Rydberg blockade $B$. The inhomogeneous laser intensity results in inhomogeneous light shifts, which induce detrimental detuning errors. (b) Protocols introduced in this work. Top: robust gates with composite (shown) or simultaneous (not shown) Rydberg and single-qubit control. Middle: pseudo-robust gates, where known detuning errors are mapped to correctable single-qubit errors. Bottom: fly-by gate, tailored to a fixed, calibrated differential Doppler shift. (c) Infidelity under static amplitude and detuning errors for the time-optimal gate~\cite{janduraTimeOptimalTwoThreeQubit2022,paganoTO}, the amplitude-robust gate~\cite{janduraOptimizingRydbergGates2023}, and the (composite) detuning-robust, amplitude- and detuning-robust, and Stark-robust protocols introduced in this work. The Stark-robust results are for Stark correlation $\zeta \approx 2$, which quantifies the detuning error introduced by amplitude errors. (d) Schematic of the regimes of intensity inhomogeneity in which the time-optimal and the detuning-robust (or pseudo-robust) protocols perform best, for two-photon Rydberg excitation.}
\end{figure}

In this work, we show that this no-go theorem can be circumvented by considering a more general, but experimentally relevant, class of Rydberg gates where both qubit states $\ket{0}$ and $\ket{1}$ take part in the gate protocol (Sec.~\ref{sec:physical-setup}). This principle can be applied in two different ways, which we lay out and characterize (Fig.~\ref{fig:overview}):
\begin{itemize}
    \item Detuning-robust and Stark-robust gates, where the $\ket{0} \leftrightarrow \ket{1}$ and $\ket{1} \leftrightarrow \ket{r}$ transitions are both driven -- either sequentially in a composite gate, or simultaneously in a co-driven gate. Equivalently, this setup is achievable by using a two-tone Rydberg scheme.
    \item \emph{Pseudo-robust} gates, requiring only the Rydberg drive, where a known detuning error is converted into a known single-qubit phase error, which can be either compensated by a symmetric single-qubit phase gate or echoed out when the quantum circuit allows for it.
\end{itemize}

The protocols are first outlined in the context of pure detuning errors in Sec.~\ref{sec:pure-det}, before being generalized to families of Stark-robust and Stark pseudo-robust protocols -- alongside an analysis of the potential improvement over the time-optimal protocol in a semi-realistic scenario in Sec.~\ref{sec:finite-stark}. Here, \emph{Stark-robust} refers to robustness against the correlated amplitude--detuning error channel generated by intensity fluctuations through light shifts.

Finally, our work addresses a separate but related question, relevant to recent proposals of constant-velocity neutral-atom architectures~\cite{lib2026velocityenabledquantumcomputingneutral,Dudinets2026AlltoallConnectivityRydbergatombased}: is there a protocol for a ``fly-by'' Rydberg entangling gate, where the two atoms experience different detunings due to sizable Doppler shifts? We show that, while this setup induces additional bright-dark state coupling in the Rydberg manifold, there exists a simple entangling gate protocol that requires only the $\ket{1}\leftrightarrow\ket{r}$ Rydberg drive (Sec.~\ref{sec:flyby}).

For these three protocols, we propose characterizations of the fidelity sensitivity to unitary and dissipative error sources (Table~\ref{tab:comparison}, Appendix~\ref{app:response}). We use them to delineate, in terms of Rydberg decay rate, intensity inhomogeneity and light-shift strength, the regime in which the new protocols outperform the time-optimal gate, and find that it includes conditions that are already relevant for state-of-the-art two-photon experiments: in a semi-realistic model at a 1\% intensity RMS and typical Rydberg decay rates, the robust protocols reduce the gate error about fivefold compared with the commonly used time-optimal gate~\cite{janduraTimeOptimalTwoThreeQubit2022,paganoTO}.

\section{\label{sec:physical-setup}Physical setup and background}

\subsection{Rydberg entangling gates}

We first consider the conventional framework of symmetric Rydberg excitation with infinite blockade, which involves two neighboring atoms $i=1,2$ with three levels each, $\ket{0}_i$, $\ket{1}_i$, and $\ket{r}_i$. $\ket{1}$ and $\ket{r}$ are coupled with complex Rabi frequency $\Omega = |\Omega| e^{i \phi}$. In this scenario, the Hamiltonian is effectively block-diagonal, and it is sufficient to consider blocks $\ket{01} \overset{\Omega}{\leftrightarrow} \ket{0r}$ and $\ket{11} \overset{\sqrt{2} \Omega}{\leftrightarrow} \ket{W}$ where $\ket{W} = (\ket{1r} + \ket{r1}) / \sqrt{2}$ is the Rydberg bright state. Thanks to the $\sqrt{2}$ enhancement factor in the second block, it is possible to design different, closed evolution trajectories where $\ket{01}$ and $\ket{11}$ acquire different phases $\theta_{01}$ and $\theta_{11}$~\cite{levineParallelImplementationHighFidelity2019} ($\theta_{10} = \theta_{01}$ by exchange symmetry). When $\theta_{11} - \theta_{01} - \theta_{10} = \theta_{11} - 2\theta_{01} = \pi$, the resulting unitary (restricted to the qubit subspace) can be factored as $R_{\theta_{01}}^{\otimes 2} \, CZ$ where $R_{\theta} \equiv \mathrm{diag}(1,e^{i\theta})$ is a single-qubit phase gate, corresponding to a global Bloch sphere rotation for the qubit subspace, and $CZ$ is the entangling gate $\mathrm{diag}(1,1,1,-1)$. In this work, we mostly assume that the Rabi frequency is kept at its maximum value, i.e., $|\Omega|=\Omega_{\mathrm{max}}$, so that the only control parameter is the phase $\phi(t)$.

The closed trajectories can be designed as piecewise constant with different modulation phases $\phi$ for each piece~\cite{levineParallelImplementationHighFidelity2019,fromonteilProtocolsRydbergEntangling2023}, or with a continuously modulated phase found via optimal control. The shortest protocol of the latter kind, the time-optimal gate~\cite{janduraTimeOptimalTwoThreeQubit2022,paganoTO}, and variations thereof are currently the main protocols in experimental use.

\subsection{Stark-correlated errors}
\label{sec:stark_corr}

Although amplitude and detuning errors are often treated as independent error channels, they need not be independent in a physical implementation. In particular, laser-intensity fluctuations modify both the Rabi coupling and the differential light shift of an optically driven transition, thereby producing correlated amplitude and detuning errors. Following Ref.~\cite{janduraOptimizingRydbergGates2023}, we characterize this correlation by the dimensionless Stark correlation
\begin{equation}
    \zeta_{ij} \equiv \frac{2\Delta_{\mathrm{LS}}^{(ij)}}{|\Omega|},
    \label{eq:zeta_main}
\end{equation}
where $\Delta_{\mathrm{LS}}^{(ij)}$ is the differential light shift between $|1\rangle$ and $|r\rangle$ induced by the laser driving
$i\leftrightarrow j$. A fractional Rabi-amplitude error $\epsilon$ caused by an intensity fluctuation is then accompanied by a detuning error
\begin{equation}
    \delta = \zeta_{ij} |\Omega| \epsilon .
\end{equation}
Thus, $\zeta_{ij}$ quantifies how strongly intensity fluctuations are converted into detuning fluctuations relative to the driven coupling, while its sign specifies the direction of the induced resonance shift.

While Stark-correlated errors are a general feature of optically driven transitions and also arise in single-photon Rydberg schemes, they are particularly important for two-photon excitation schemes commonly used in alkali atoms, where the Stark correlation can become large ($|\zeta|>1$). For a ladder transition through an intermediate state $|e\rangle$ with single-photon detuning $\Delta$, adiabatic elimination yields the effective Rydberg coupling $\Omega=\Omega_{1e}\Omega_{er}/(2\Delta)$. The dependence of the resulting Stark correlations on the atomic parameters and laser configuration is discussed in detail in
Appendix~\ref{app:error_scalings}.

For balanced one-photon Rabi frequencies, $\beta\equiv |\Omega_{1e}/\Omega_{er}|=1$, the lower-leg Stark correlation $\zeta_{1e}$ depends only weakly on $|\Delta|$ over the experimentally relevant regime, whereas the magnitude of the upper-leg correlation grows asymptotically as
\begin{equation}
    |\zeta_{er}| \propto n^3 |\Delta| ,
    \label{eq:zeta_upper_main}
\end{equation}
with the sign determined by the corresponding differential polarizability, where \(n\) is the (effective) principal quantum number of the Rydberg state (Appendix~\ref{app:error_scalings}).

This scaling introduces a trade-off with the usual strategies for suppressing dissipative errors. Increasing $|\Delta|$ suppresses intermediate-state scattering, while increasing $n$ reduces Rydberg-state decay. Both choices, however, tend to increase the magnitude of the upper-leg Stark correlation and therefore the sensitivity to upper-leg intensity fluctuations. Conversely, reducing $|\zeta_{er}|$ favors smaller $|\Delta|$ and smaller $n$, in competition with these dissipative-error considerations.

This trade-off is already relevant in current experiments. For example, the parameters of Ref.~\cite{evered2026highfidelityentanglinggatesnonlocal} correspond to $|\zeta_{er}|\simeq 6$. A detailed treatment of the two-photon excitation scheme, including intermediate-state elimination, the scaling of the relevant dissipative, frequency-noise, and intensity-noise errors with $\Omega$, $\Delta$, $n$, and $\beta$, and representative Stark correlations for common Cs and Rb implementations, is given in Appendix~\ref{app:error_scalings}.

\subsection{Formalism}

For the sake of simplicity, we restrict our investigation of robust gate protocols to the symmetric Rydberg Hamiltonian with infinite blockade, and we use adiabatic elimination to disregard $\ket{e}$. We also consider symmetric detuning and amplitude errors, which is justified when considering errors caused by imperfect beam shaping or intensity fluctuations since these affect neighboring, $\upmu$m-separated atoms roughly identically.

The Rydberg Hamiltonian is block-diagonal, $H^R = H_{00}^R \oplus H_{01}^R \oplus H_{10}^R \oplus H_{11}^R$, with blocks

\begin{subequations}\label{eq:HR}
\begin{eqnarray}
H_{00}^R &=& 0,\\
H_{01}^R &=& \frac{(1+\epsilon)\Omega}{2} \ket{01}\bra{0r} + \mathrm{h.c.} + \delta \ket{0r}\bra{0r},\\
H_{10}^R &=& \frac{(1+\epsilon)\Omega}{2} \ket{10}\bra{r0} + \mathrm{h.c.} + \delta \ket{r0}\bra{r0}, \\
H_{11}^R &=& \frac{(1+\epsilon)\Omega}{\sqrt{2}} \ket{11}\bra{W} + \mathrm{h.c.} + \delta \ket{W}\bra{W},
\end{eqnarray}
\end{subequations}
where $\epsilon$ is a fractional amplitude error and $\delta$ is a detuning error, i.e., a shift of the two-photon resonance. The doubly excited state $\ket{rr}$ is excluded by the infinite blockade, and the Rydberg dark state $\ket{D} = (\ket{1r} - \ket{r1})/\sqrt{2}$ is decoupled. We denote by $P_r = \ket{0r}\bra{0r} + \ket{r0}\bra{r0} + \ket{W}\bra{W}$ the projector onto the Rydberg states of this symmetric subspace, so that $\partial H^R / \partial \delta = P_r$.

We also consider a single-qubit drive $h^1 = \frac{\Omega_1}{2} \ket{0}\bra{1} + \mathrm{h.c.}$ with $\Omega_1 = |\Omega_1| e^{i\psi}$, resulting in a Hamiltonian

\begin{equation}
    H^1 = I \otimes h^1 + h^1 \otimes I.
\end{equation}

We quantify the gate performance using the average subspace gate fidelity, which, for a target unitary $U_t$, a realized unitary $U$, a computational subspace projector $P = \ket{00}\bra{00} + \ket{10}\bra{10} + \ket{01}\bra{01} + \ket{11}\bra{11}$, and computational subspace dimension $D=\mathrm{tr}(P) = 4$, is given by~\cite{Pedersen2007FidelityQuantumOperations}
\begin{equation}
    F(U,U_t)=\frac{\mathrm{tr}(M M^{\dagger}) + |\mathrm{tr}(M)|^2}{D(D+1)}, \qquad M = P U_t^{\dagger} U P.\label{eq:fidelity}
\end{equation}

We say that a protocol based on phase modulation $\phi(t)\in \mathbb{R}$ ($t\in[0,T]$) produces a $CZ_{\theta}$ entangling gate if the time evolution of $H^R$ (with $\epsilon=0$ and $\delta=0$) with these control parameters results in a final unitary $U(T)=U_t = R_{\theta}^{\otimes 2} CZ$.

Throughout this work, we consider amplitude ($\epsilon$) and detuning ($\delta$) unitary errors that are quasi-static, i.e. that are time-independent for a given realization of the unitary $U$: they can be either static, and thus can be known by calibration, or stochastic. We usually assume that they follow Gaussian distributions with standard deviations $\sigma_{\epsilon}$ and $\sigma_{\delta}$. We denote by $U(T,\epsilon,\delta)$ the unitary resulting from the time evolution of $H^R$ with errors $\epsilon$ and $\delta$.

We say that this protocol is fully robust to quasi-static amplitude errors when
\begin{equation}
\left.\frac{\partial^2 F(U(T,\epsilon,0),U_t)}{\partial \epsilon^2}\right|_{\epsilon=0} = 0,\label{eq:amp-robust}
\end{equation}
and to quasi-static detuning errors when
\begin{equation}
\left.\frac{\partial^2 F(U(T,0,\delta),U_t)}{\partial \delta^2}\right|_{\delta=0} = 0.
\end{equation}
We quantify the residual sensitivity of a protocol by $S_\epsilon \equiv -\frac{1}{2}\,\partial^2 F/\partial \epsilon^2$ and $S_\delta \equiv -\frac{1}{2}\,\partial^2 F/\partial \delta^2$.

When the detuning error is correlated with the amplitude error as $\delta = \zeta |\Omega| \epsilon$, Eq.~\ref{eq:amp-robust} expresses the condition for robustness to this correlated Stark-shift channel (with $U(T,\epsilon,0)$ replaced by $U(T,\epsilon,\zeta|\Omega|\epsilon)$).

\subsection{Detuning-error robustness}

\subsubsection{No-go theorem for detuning-robust conventional gates}

The time-optimal gate~\cite{janduraTimeOptimalTwoThreeQubit2022,paganoTO} provides the shortest possible protocol that achieves an entangling gate and has error sensitivities $\frac{\partial^2 F}{\partial \epsilon^2} \approx -8.0$ and $\frac{\partial^2 F}{\partial \delta^2} \approx -5.7/|\Omega|^2$ (Table~\ref{tab:comparison}).

Ref.~\cite{janduraOptimizingRydbergGates2023} derives the shortest possible protocol that fulfills Eq.~\ref{eq:amp-robust} (an amplitude-robust gate), finding a duration $T \approx 14.32/|\Omega|$, i.e., 88\% longer than the time-optimal gate.
Ref.~\cite{janduraOptimizingRydbergGates2023} also demonstrates that a conventional Rydberg protocol (where only $\ket{1}$ is coupled to $\ket{r}$) can never achieve robustness against detuning errors. It is helpful to summarize the argument as follows: the first-order perturbation of the gate unitary is given by
\begin{eqnarray}
\left.\frac{\partial U(T,0,\delta)}{\partial \delta}\right|_{\delta=0} &=& -i U(T) N_r\\
\text{with } N_r &=&\int_0^T{U(t)^{\dagger} P_r U(t) dt},
\end{eqnarray}
where $P_r = \partial H^R / \partial \delta$ is the detuning error operator. $N_r$ is a Hermitian operator whose diagonal elements,
\begin{equation}
N_q \equiv \bra{q} N_r \ket{q} = \int_0^T \bra{\psi_q(t)} P_r \ket{\psi_q(t)} dt,\label{eq:Nq}
\end{equation}
are the time-integrated Rydberg populations along the trajectories $\ket{\psi_q(t)} = U(t)\ket{q}$ starting from the computational basis states $\ket{q}$. We refer to $N_q$ as the \emph{Rydberg dwell time} of $\ket{q}$; in particular, the phase of each computational state drifts at first order as $\theta_q \to \theta_q - N_q \delta$. Crucially, $P_r$ (and thus $N_r$) is positive semi-definite. The gate is insensitive to $\delta$ at first order iff there exists a real $\kappa$ such that
\begin{equation}
\left.\frac{\partial U(T,0,\delta)}{\partial \delta}\right|_{\delta=0}P = -i\kappa U(T)P,\label{eq:first-order-robust}
\end{equation}
i.e., iff the restriction of $U(T)$ to the computational subspace is unchanged up to a global phase. This requires
\begin{equation}
N_r P = \kappa P,
\end{equation}
i.e., the Rydberg dwell time must be the same for every computational basis state, and $N_r$ must have no matrix elements between computational and Rydberg states. Since $\ket{00}$ is not coupled to the drive ($H^R_{00}=0$), $N_{00}=0$, whereas $N_q>0$ for the other basis states of any entangling protocol, so this condition can never be satisfied.

\subsubsection{Speed limit for Stark-robust conventional gates}\label{sec:qsl-mt}

This argument provides two routes to circumvent the no-go theorem:

\begin{enumerate}[label=(\roman*)]
    \item modify the error operator such that it is not positive semi-definite; \label{itm:non-semidef}
    \item involve $\ket{00}$ in the gate protocol.
\end{enumerate}

\ref{itm:non-semidef} can be accomplished by considering a finite Stark correlation, $0 < |\zeta| < \infty$, between the amplitude and detuning errors~\cite{janduraOptimizingRydbergGates2023}. The resulting error operator is indeed not positive semi-definite; however, as we now show, the gate duration then scales poorly with $\zeta$. As such, the following results extend the strict no-go theorem, which is recovered in the pure-detuning limit $\zeta \to \infty$, to finite $\zeta$ in terms of the required gate durations with conventional Rydberg protocols.

To this end, consider the block-diagonal error operator for computational basis state indices $q\in \{01,10,11\}$, with maximal drive strength $\Omega = e^{i \phi} |\Omega|$,
\begin{equation}
\frac{\partial H^R_q}{\partial \epsilon} = |\Omega|
\begin{pmatrix}
0 & \chi_q\frac{e^{i \phi}}{2}\\
\chi_q \frac{e^{-i\phi}}{2} & \zeta
\end{pmatrix},
\end{equation}
where $\chi_q$ is the block's enhancement factor (1 for $01$ and $10$, $\sqrt{2}$ for $11$). Since $\ket{00}$ is still dark in the protocol, the robustness criterion requires
\begin{equation}
\bra{q}\left(\int_0^T{U(t)^{\dagger} \frac{\partial H^R_q}{\partial \epsilon} U(t) dt} \right)\ket{q} =0.
\end{equation}

Using an analogous argument to typical derivations of quantum speed limits (via the Cauchy-Schwarz inequality on the instantaneous Rydberg population $|\langle r(q) | \psi_q(t)\rangle|^2 = 1-|\langle q|\psi_q(t)\rangle|^2$, where $\ket{r(q)}$ is the Rydberg state coupled to $\ket{q}$), we show in Appendix~\ref{app:qsl} that the duration of a Stark-robust gate with Stark correlation $\zeta$ is lower-bounded by
\begin{equation}
T \geq N_q \max\left(\frac{2|\zeta|}{\chi_q}, \frac{\zeta^2}{\chi_q^2}\right),\label{eq:qsl-bound}
\end{equation}
so that the minimum gate duration grows quadratically with $\zeta$ for $|\zeta| \gtrsim 2\chi_q$.

As an example, using the Rydberg populations from the time-optimal gate (whose subspace-averaged Rydberg population is within 1\% of the minimum), this bound would require $T \gtrsim 144/|\Omega|$ for $|\zeta|=6$, the binding constraint arising from the $\ket{01}$ block ($T \geq \zeta^2 N_{01}$ with $N_{01} \approx 4.0/|\Omega|$), which would be prohibitively costly for the gate's error budget. This estimate is loose, in practice, because this gate would certainly require considerably higher Rydberg population than the time-optimal gate. As an example, Ref.~\cite{janduraOptimizingRydbergGates2023} found that numerical optimization of a Stark-robust gate fails for $\zeta \gtrsim 2$.

On the other hand, the relevant regime to mitigate the extreme sensitivity of the Rydberg two-photon transition to light shifts involves $|\zeta| \gtrsim 5$, which further motivates the need for new classes of entangling gate protocols that enable detuning robustness.

\section{Detuning-robust and pseudo-robust protocols}\label{sec:pure-det}

\subsection{Detuning-robust protocols}

\subsubsection{Composite gate}

\begin{figure}
\centering
    \includegraphics[width=86mm]{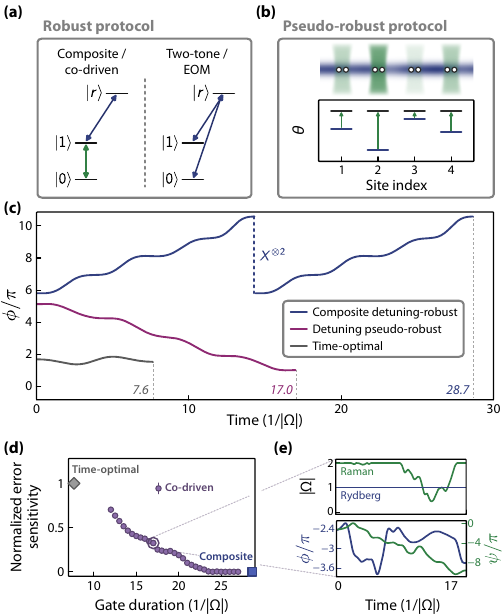}
    \caption{\label{fig:pure-det}\textbf{Detuning-robust and pseudo-robust protocols.} (a) Level diagram for the robust protocols, realized by driving both the qubit and Rydberg transitions, or by using a two-tone Rydberg scheme. (b) Schematic of the pseudo-robust protocol, where detuning errors are not suppressed but converted into correctable single-qubit errors. (c) Phase modulation waveforms and gate durations for the time-optimal protocol~\cite{janduraTimeOptimalTwoThreeQubit2022,paganoTO}, the composite detuning-robust gate, and the detuning pseudo-robust gate. (d) Required gate duration to achieve a given detuning error sensitivity reduction with the co-driven protocol, compared with the time-optimal gate and the composite detuning-robust gate. (e) Amplitude and phase waveforms for the circled gate in (d), corresponding to 3-fold reduced sensitivity compared with the time-optimal gate.}
\end{figure}

We first describe protocols robust to pure detuning errors in this section, followed by Stark-robust protocols in the next section. Generally, unitary errors on conventional, symmetric, Rydberg $CZ$ gates can be of three types:

\begin{enumerate}[label=(\roman*)]
    \item leftover population in Rydberg state (leakage);
    \item single-qubit errors $I Z+Z I$;\label{itm:1q-err}
    \item entangling phase errors $Z Z$.\label{itm:2q-error}
\end{enumerate}

In the previous section we showed that conventional Rydberg pulses must result in errors of type \ref{itm:1q-err} or \ref{itm:2q-error} -- in addition to possible leakage errors. Both protocol classes of this section rest on the same first-order condition: since the phase of computational basis state $\ket{q}$ drifts as $\theta_q \to \theta_q - N_q \delta$, choosing the Rydberg dwell times to satisfy $N_{11} = N_{01} + N_{10}$ removes the entangling ($ZZ$) error, leaving (absent leakage) only single-qubit phase drifts -- the most that can be achieved while $\ket{00}$ remains dark, since full robustness would require all four $N_q$ to be equal. The two classes differ in how they use this condition: the composite robust gate cancels the remaining drifts with $X^{\otimes 2}$ echoes -- thereby involving $\ket{00}$ and evading the no-go theorem -- while the pseudo-robust gate keeps the conventional drive and corrects the known drifts with a local phase gate.

Full robustness thus requires involving both qubit states in the dynamics, by driving both the Rydberg and qubit transitions or, equivalently, with a two-tone Rydberg drive (Fig.~\ref{fig:pure-det}(a)). The simplest such protocol is the composite gate, where $X^{\otimes 2}$ echoes are inserted in the gate sequence: 
\begin{equation}
U_c = X^{\otimes 2} VX^{\otimes 2} V,
\end{equation}
where $V=R_{\theta}^{\otimes 2} C_{\pi/2}$ is obtained as a conventional Rydberg pulse. $C_{\pi/2}$ denotes a controlled-$R_{\pi/2}$ gate, given in the computational basis as $\mathrm{diag}(1,1,1,i)$. Since $X^{\otimes 2}$ exchanges $\ket{00}\leftrightarrow\ket{11}$ and $\ket{01}\leftrightarrow\ket{10}$, the phases of the two halves add up as $\theta_q + \theta_{\bar{q}}$, and $U_c$ realizes the entangling gate
\begin{equation}
    U_c = i e^{2i\theta} \mathrm{diag}(1,-i,-i,1) = i e^{2i\theta} R_{-\pi/2}^{\otimes 2} CZ.
\end{equation}

The robustness condition yields several requirements for $\left.\partial V/\partial \delta\right|_{\delta=0}$.

\begin{itemize}
    \item $V(\delta)$ must not lead to leakage out of the computational states. Consider, for instance, leftover population from $V\ket{01}$ in $\ket{0r}$, which becomes $\ket{1r} = (\ket{W} + \ket{D})/\sqrt{2}$ after the action of $X^{\otimes 2}$: since $H^R\ket{D} = 0$, the $\ket{D}$ component survives the second half-gate and results in leftover population in the Rydberg states after $U_c(\delta)$. Likewise, leftover population from $V\ket{11}$ in $\ket{W}$ becomes $(\ket{0r}+\ket{r0})/\sqrt{2}$, which eventually becomes leftover population in $\ket{W}$.

    \item{In the absence of leakage, the first-order error generator of $V$ is diagonal and symmetric and thus can be expanded in the $\{I,I Z +Z I,Z Z\}$ basis. $IZ+ZI$ anticommutes with $X^{\otimes 2}$, so these errors are cancelled by the echo. However, $ZZ$ errors commute with $X^{\otimes 2}$, so they do propagate into the final error operator. Therefore the error generator of $V$ must not contain $ZZ$ terms. Using $\theta_q \to \theta_q - N_q \delta$ and $N_{00} = 0$, this condition can also be expressed as
    \begin{equation}
    N_{11} = N_{01} + N_{10} = 2N_{01},\label{eq:dwell-balance}
    \end{equation}
    which we refer to as the \emph{Rydberg-dwell balance}.}
\end{itemize}

In summary, the underlying condition for a fully robust composite gate is that half-gates $V$ only have type-\ref{itm:1q-err} errors. This setup bears close resemblance to previous proposals cancelling Doppler shifts via a tweezer-induced detuning reversal between two $C_{\pi/2}$ pulses (e.g., Protocol II.a of Ref.~\cite{fromonteilProtocolsRydbergEntangling2023}, of duration $T \approx 21.45/|\Omega|$), with an important nuance: there, $ZZ$ terms are also cancelled by the detuning reversal; hence, the conditions on the half-gate are more stringent for the composite gate in our case.

\subsubsection{Numerical optimization of the composite gate}

Following the principles of GRadient Ascent Pulse Engineering~\cite{Khaneja2005OptimalControlCoupled} (GRAPE), the half-gate phase modulation $\phi(t) \ (t\in [0,T/2])$ is written as a piecewise constant function and optimized with a cost function (alongside the single-qubit phase $\theta$) that adds to the infidelity a penalty enforcing precisely the two requirements above -- no first-order leakage, and the Rydberg-dwell balance (Eq.~\ref{eq:half-gate-req}, Appendix~\ref{app:numerics}).

The optimization procedure yields, at $T \geq 28.68/|\Omega|$, the \emph{composite detuning-robust protocol} (Fig.~\ref{fig:sm-convergence}). The corresponding modulation waveform is shown in Fig.~\ref{fig:pure-det}(c). While this protocol is $3.8\times$ longer than the time-optimal gate protocol ($T= 7.61/|\Omega|$), its sensitivity to quasi-static amplitude errors is only $1.57\times$ higher, and its total Rydberg dwell time is $2.18\times$ larger (Table~\ref{tab:comparison}).

In Appendix~\ref{app:more-gates}, we further show the existence of a composite gate robust to both amplitude and detuning errors, requiring a total time $T\gtrsim 46.5/|\Omega|$.

\subsubsection{Co-driven gates for partial robustness}

Full detuning robustness, i.e., $S_\delta = 0$, is not required to achieve a practically relevant gain for high-fidelity entangling gates. In alkali atoms, for which the adverse scaling of detuning errors in two-photon Rydberg transitions is most deleterious, single-qubit Rabi frequencies of the same order of magnitude as Rydberg Rabi frequencies are achievable via optical Raman control~\cite{levineDispersiveOpticalSystems2022}. This makes it possible to drive the qubit and Rydberg transitions simultaneously, in what we refer to as a \emph{co-driven} gate.

We work under the assumption that the maximum single-qubit Rabi frequency is twice the Rydberg Rabi frequency ($\Omega_{1,\mathrm{max}} = 2\Omega_{\mathrm{max}}$). With a second, amplitude-bounded drive present, keeping the Rydberg amplitude maximal is no longer obviously optimal; however, we find that allowing for a variable Rydberg amplitude does not enable better protocols, and therefore retain the standard assumption $|\Omega| = \Omega_{\mathrm{max}}$.

Due to the single-qubit drive, the Hamiltonian is no longer block-diagonal, and the unitary needs to be optimized on the entire blockaded qubit--Rydberg subspace spanned by 8 states ($\{\ket{0},\ket{1},\ket{r}\}^{\otimes 2}\backslash \{\ket{rr}\}$).

Rather than constraining the optimization target to be a $CZ_{\theta}$ gate, we perform a numerical search to approximate any entangling gate, expressed as a $CZ$ gate with a general, single-qubit symmetric gauge
\begin{equation}
    U_t = R(u,v,w)^{\otimes 2} \cdot CZ \cdot R(0,v',w')^{\otimes 2},\label{eq:local-target}
\end{equation}
where $R(u,v,w) = R_z(u) R_x(v) R_z(w)$ denotes a single-qubit rotation parameterized by Euler angles. (The $z$-rotation adjacent to $CZ$ is omitted from the right factor, since it commutes with $CZ$ and can be absorbed into the left factor.)
The unitary's functional parameters (Rydberg phase $\phi(t)$, qubit-drive phase $\psi(t)$, qubit-drive amplitude $|\Omega_1|(t)$) are jointly optimized with the Euler angles, using a tunable weight for the robustness penalty which trades off residual sensitivity against gate duration. Only gates where $F>1-10^{-5}$ are selected.

The required gate times to achieve a given level of robustness compared to the time-optimal gate are presented in Fig.~\ref{fig:pure-det}(d). Additionally, the phase and amplitude modulations that yield a gate with $3\times$ error suppression are shown in Fig.~\ref{fig:pure-det}(e). Interestingly, while this protocol enables partial detuning sensitivity suppression at a shorter gate duration than the composite gate, its gate duration converges to the composite gate's duration as the required sensitivity is lowered to 0. This indicates that the composite protocol is optimal within the broader class of fully robust protocols involving both qubit states.

Two-tone Rydberg driving, i.e., separate $\ket{1}\overset{e^{i\phi_1}}{\leftrightarrow} \ket{r}$ and $\ket{0}\overset{e^{i\phi_0}}{\leftrightarrow} \ket{r}$ complex Rabi frequencies, provides another experimentally relevant way to involve both qubit states, and is in fact rigorously equivalent to the co-driven scheme via a time-dependent basis change. In this context the bound on the single-qubit Rabi frequency becomes a limit on the slew rate $|\dot{\phi}_1 - \dot{\phi}_0|$. The same equivalence applies to the modulation of a single Rydberg laser with an electro-optic modulator (EOM) at the qubit frequency (Appendix~\ref{app:two-tone}).

\subsection{Detuning pseudo-robust protocols}\label{sec:pseudo}

\subsubsection{Setup and optimization}

The longer gate duration and increased error sensitivities of the robust protocol motivate the introduction of a different class of entangling protocols, which we call \emph{pseudo-robust}: since detuning errors in the two-photon transition setup are primarily caused by static beam shaping inhomogeneities that can be experimentally measured (for instance, using Ramsey interferometry) for each gate site, it is sufficient to \emph{convert} detuning errors into single-qubit phase errors that can be corrected. The protocol is summarized in Fig.~\ref{fig:pure-det}(b).

In hyperfine qubits, local single-qubit phase gates can be straightforwardly implemented by tuning a laser close to an electronic transition between the electronic state hosting the qubit and another electronic state, resulting in a differential light shift on the qubit. Since the acquired single-qubit phase is proportional to the local laser intensity, a local phase can be implemented using, e.g., a spatial light modulator to imprint the required static intensity pattern. Because the correction is already a second-order effect, under- or over-rotations on the correction are a higher-order concern, and the scattering cost of a light-shift correction is negligible for realistic detuning noise (Appendix~\ref{app:correction-cost}). 

Furthermore, the correction does not require narrow-linewidth lasers, high power or a specific polarization. Finally, a quantum circuit can be amenable to an $X$ echo between consecutive entangling gates (e.g., the echo-RB protocol of Ref.~\cite{everedHighfidelityParallelEntangling2023}), in which case no correction is required.

The requirements for pseudo-robustness correspond exactly to the composite robust protocol's requirements for the half-gate $V$ -- no leakage and the Rydberg-dwell balance of Eq.~\ref{eq:dwell-balance} -- i.e., that a detuning error only results in single-qubit $IZ+ZI$ errors. Formally, this is expressed by simply replacing the $R_{\theta}^{\otimes 2} C_{\pi/2}$ target in Eq.~\ref{eq:half-gate-req} with $R_{\theta}^{\otimes 2} CZ$.

Using the same optimization techniques, we find that a pseudo-robust $CZ$ gate is achievable for $T\gtrsim 17.04/|\Omega|$, which is 41\% shorter than the fully robust gate, and we present the corresponding phase modulation waveform in Fig.~\ref{fig:pure-det}(c). This protocol's duration and integrated Rydberg population are close to those of the amplitude-robust gate~\cite{janduraOptimizingRydbergGates2023} ($T\approx 14.3/|\Omega|$), which has been recently experimentally realized~\cite{liu2026highfidelityneutralatomgates} with a gate fidelity of 99.6\%.

\subsubsection{Corrected fidelity and error sensitivity}

To better understand error scaling under this protocol, it is helpful to introduce the \emph{corrected} average gate fidelity, which represents the fidelity of the unitary $U$ followed by the single-qubit correction $R_{\vartheta}^{\otimes 2}$ where $\vartheta$ is chosen to maximize the fidelity of $R_{\vartheta}^{\otimes 2} U$. For example, considering a detuning error $\delta$, the corrected fidelity can be written
\begin{equation}
\tilde{F}(U(T,\delta),U_t) = \max_{\vartheta} F\left(R_{\vartheta}^{\otimes 2} U(T,\delta),U_t\right).
\end{equation}
We can define, from the corrected fidelity, a notion of corrected sensitivity $\tilde{S}_{\cdot}$.

For a conventional $\ket{1} \leftrightarrow \ket{r}$ Rydberg gate, in the limit where $|\epsilon| \ll 1$ and $|\delta| \ll |\Omega|$, the corrected fidelity can be expressed as
\begin{align}
&\tilde{F}(U(T,\delta),R_{\theta}^{\otimes 2} CZ)\nonumber\\
&\quad = F(U(T,\delta), R_{\theta-\vartheta}^{\otimes 2} CZ),
\end{align}
where $\vartheta$ is the optimal phase correction given to leading order by
\begin{equation}
\vartheta = \frac{N_{11}-N_{00}}{2} \delta.
\end{equation}

Equivalently, for an amplitude error $\epsilon$ the correction can be written
\begin{equation}
\vartheta = \frac{D_{11}-D_{00}}{2}\epsilon,
\end{equation}
where $D_q = \int_0^T \bra{\psi_q(t)} \hat{D} \ket{\psi_q(t)} dt$ is the analog of $N_q$ for the amplitude error operator $\hat{D} = \partial H^R/\partial\epsilon = \oplus_q{\hat{D}_q}$, given by the off-diagonal part of the Rydberg Hamiltonian:
\begin{equation}
\hat{D}_q = |\Omega|
\begin{pmatrix}
0 & \chi_q\frac{e^{i \phi}}{2}\\
\chi_q \frac{e^{-i\phi}}{2} & 0
\end{pmatrix}.
\end{equation}

It is informative to distinguish between \emph{correctable} errors that can be partially or fully mitigated by locally optimizing a correction $\vartheta$, and \emph{uncorrectable} errors that cannot be, due to their stochastic nature (e.g., laser frequency noise). Even for a standard time-optimal gate, local corrections can significantly improve the corrected fidelity's sensitivity to amplitude and detuning errors. Here, the pseudo-robust protocol \emph{entirely} channels detuning errors into correctable errors, but, as is the case for other protocols, its sensitivity to correctable amplitude errors can also be decreased by including these in the correction.

More generally, the corrected sensitivity to an error $\varepsilon$ follows from the second-order expansion of $F$ in $(\varepsilon,\vartheta)$ (Appendix~\ref{app:corrected}),
\begin{equation}
\frac{\partial^2 \tilde{F}}{\partial \varepsilon^2} = \frac{\partial^2 F}{\partial \varepsilon^2} - \frac{\left(\frac{\partial^2 F}{\partial \varepsilon \partial \vartheta}\right)^2}{\frac{\partial^2 F}{\partial \vartheta^2}}.
\end{equation}
For a leakage-free gate, this gives $\tilde{S}_\delta = (N_{11} + N_{00} - 2N_{01})^2/20$, which vanishes exactly under the Rydberg-dwell balance.

We summarize in Table~\ref{tab:comparison} the regular and corrected fidelity sensitivities for the protocols mentioned in this work. We also find that there exists a detuning and amplitude pseudo-robust gate for $T\gtrsim 24.27/|\Omega|$. Incidentally, the protocol turns out to be not only amplitude pseudo-robust, but also amplitude-robust (up to a small error).

\section{Protocols for finite Stark correlation}\label{sec:finite-stark}

\begin{figure}
\centering
    \includegraphics[width=84mm]{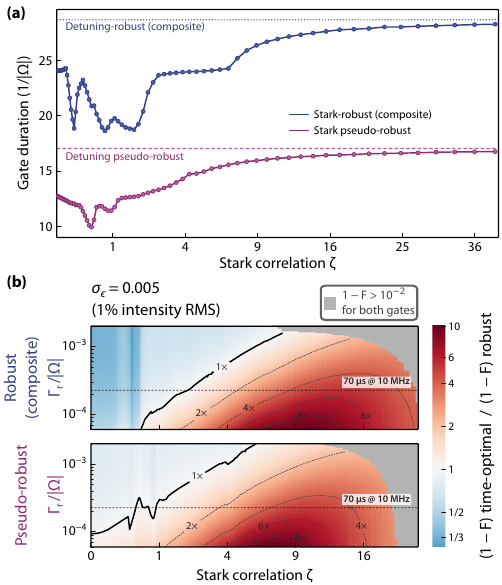}
    \caption{\label{fig:stark}\textbf{Stark-robust and pseudo-robust protocols.} (a) Minimal gate duration for composite Stark-robust and Stark pseudo-robust protocols with correlation $\zeta$. At $\zeta \to \infty$ the protocols converge to their detuning-robust versions. (b) Ratio of expected infidelity from the time-optimal gate to that of the robust and pseudo-robust protocol depending on $\zeta$ and the normalized Rydberg decay rate $\Gamma_r/|\Omega|$, for a Gaussian-distributed static relative intensity error with 1\% RMS. Regions where none of the three protocols achieves $F>1-10^{-2}$ are greyed out.}
\end{figure}

\subsection{Finite Stark correlation}

The previous section demonstrates how the robust and pseudo-robust constructions circumvent the no-go theorem that prohibits detuning-robust Rydberg gates under the conventional protocol. This pure detuning-robust case can be thought of as the $\zeta\to \infty$ limit of Stark-robust gates, while previous results for amplitude-robust gates~\cite{janduraOptimizingRydbergGates2023} correspond to the $\zeta\to 0$ limit (see Fig.~\ref{fig:overview}(c)).

We now lay out how the same constructions as in the previous section can be applied to the more general case of finite Stark correlation $0<|\zeta|<\infty$, and bypass the quantum speed limit for Stark-robust gates from Sec.~\ref{sec:qsl-mt}. Because exchanging $\phi(t) \to -\phi(t)$, $\theta \to -\theta$, $\vartheta \to -\vartheta$ (for the pseudo-robust case), and $\zeta \to -\zeta$ realizes an involution on robust (or pseudo-robust) $CZ_{\theta}$ protocols, it is sufficient to analyze the finite-$\zeta$ protocols in the case of positive Stark correlation.

The numerical optimizations (with cost function from Eq.~\ref{eq:half-gate-req}) for $\zeta\to\infty$ can straightforwardly be adapted to the finite $\zeta$ case: the optimized quantity becomes $S_{\epsilon}/(1+\zeta^2)$ (or $\tilde{S}_{\epsilon}/(1+\zeta^2)$ for the pseudo-robust case) with coupling $\delta = \zeta |\Omega| \epsilon$, where the normalization $1/(1+\zeta^2)$ helps preserve the cost function's well-behavedness.

The required gate durations to achieve robustness or pseudo-robustness to the $\zeta$-correlated error are shown in Fig.~\ref{fig:stark}(a). As expected, the protocol durations converge to their detuning-only values as $\zeta\to\infty$ ($T \to 28.68/|\Omega|$ and $17.04/|\Omega|$, respectively), and are significantly lower at finite $\zeta$, reaching minima of $T \approx 18.6/|\Omega|$ at $\zeta \approx 0.8$ (robust) and $T \approx 9.9/|\Omega|$ at $\zeta \approx 0.5$ (pseudo-robust) -- which matches the intuition that finite correlations yield more accessible robustness than detuning-only errors.

\subsection{Protocol performance with Rydberg decay at finite Stark correlation}

\begin{table*}
\caption{\label{tab:comparison}\textbf{Comparison of the main protocols listed in this work}, including the gate duration and the Rydberg dwell time averaged over the computational basis states (both in units of $1/|\Omega|$), and the amplitude and detuning sensitivities $S_\epsilon$ and $S_\delta$. \emph{Correctable} columns for error sensitivities refer to the corrected fidelity, under the assumption that these errors are suppressed as much as possible using local single-qubit phase corrections. Normalized error sensitivities lower than $10^{-3}$ are approximated to 0.}
\begin{ruledtabular}
\begin{tabular}{ccccccc}
\textbf{Protocol} & $\bm{|\Omega|T}$ & $\bm{|\Omega|\langle N_r\rangle}$ & \multicolumn{2}{c}{$\bm{S_\epsilon}$} & \multicolumn{2}{c}{$\bm{|\Omega|^2 S_\delta}$}\\
 & & & \textbf{Uncorrectable} & \textbf{Correctable} &  \textbf{Uncorrectable} & \textbf{Correctable}\\ \hline
Time-optimal~\cite{janduraTimeOptimalTwoThreeQubit2022,paganoTO} & 7.61 & 2.99 & 3.98 & 3.43 & 2.83 & 1.25 \\
Amplitude-robust~\cite{janduraOptimizingRydbergGates2023} & 14.32 & 4.74 & \textcolor{mygreen}{0} & \textcolor{mygreen}{0} & 6.06 & 2.02\\
Composite detuning-robust & 28.68 & 6.53 & 6.24 & 6.24 & \textcolor{mygreen}{0} & \textcolor{mygreen}{0}\\
Composite detuning- \& amplitude-robust & 46.48 & 12.61 & \textcolor{mygreen}{0} & \textcolor{mygreen}{0} & \textcolor{mygreen}{0} & \textcolor{mygreen}{0}\\
Detuning pseudo-robust & 17.04 & 4.79 & 8.83 & 7.42 & 9.17 & \textcolor{mygreen}{0}\\
Detuning \& amplitude pseudo-robust & 24.27 & 6.88 & \textcolor{mygreen}{$<10^{-2}$} & \textcolor{mygreen}{0} & 18.86 & \textcolor{mygreen}{0}\\
Fly-by ($\delta_- = |\Omega|/2$) & 12.75 & 4.70 & 4.25 & 3.89 & 9.10 & 2.75
\end{tabular}
\end{ruledtabular}
\end{table*}

Considering the protocols at finite $\zeta$ provides an indication of their experimental relevance when balancing the gained robustness with the additional errors incurred by the longer gate duration. For this purpose, we compare in Fig.~\ref{fig:stark}(b) the pseudo-robust and robust protocols to the conventional time-optimal protocol by assuming a Gaussian-distributed intensity inhomogeneity of 1\% RMS ($\sigma_\epsilon = 0.005$) at various Stark correlations $\zeta$ and normalized Rydberg decay rates $\Gamma_r/|\Omega|$. This delineates the parameter space where the new protocols provide a gain over the time-optimal gate. The assumptions and gate selection procedure are detailed in Appendix~\ref{app:fig5}.

At low intensity inhomogeneity and low Stark correlation, the gate error is dominated by the Rydberg decay, so the time-optimal protocol results in higher average fidelity. On the other hand, at higher inhomogeneity and higher Stark correlation, the robust and pseudo-robust protocols provide strong error reduction. For example, for a typical value of $\Gamma_r \approx 2.3\times10^{-4} \ |\Omega|$ (corresponding to a lifetime of 70 $\upmu$s and a Rabi frequency of 10 MHz), both the pseudo-robust and the composite protocols yield about fivefold lower error rates than the time-optimal gate at 1\% intensity RMS. For very high $\zeta \sigma_\epsilon$, quartic sensitivity reduces the gain provided by the robust and pseudo-robust protocols. As available laser power increases, lower normalized Rydberg decay rates $\Gamma_r/|\Omega|$ become accessible, which further strengthens the relevance of the robust protocols to near- and medium-term experimental progress.

Additional errors must be considered to realistically simulate the protocol performance. Nevertheless, Fig.~\ref{fig:stark}(b) is a useful proxy as it shows the trade-off between dissipative errors and intensity noise scaling. As a point of comparison, Rydberg decay represents $\sim$40\% of the simulated error budget in Ref.~\cite{evered2026highfidelityentanglinggatesnonlocal}.

\section{Fly-by $CZ$ gates}\label{sec:flyby}

\begin{figure}
\centering
    \includegraphics[width=86mm]{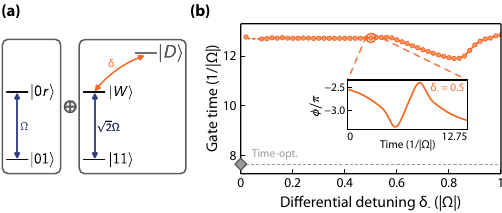}
    \caption{\label{fig:fly-by}\textbf{Fly-by gate protocol.} (a) Level diagram for the fly-by gate with differential detuning $\delta_-$. The detuning introduces a coupling to the Rydberg dark state $\ket{D}$. (b) Required gate time for a fly-by gate depending on the differential detuning $\delta_-/|\Omega|$ and example phase modulation waveform at $\delta_-/|\Omega| = 1/2$ (inset). The discontinuity between $\delta_- = 0$ and $\delta_- >0$ relaxes into a sharp but continuous gate time increase when the infidelity threshold is loosened.}
\end{figure}

\subsection{Context and setup}

The techniques developed in this work also allow us to propose a gate protocol for a \emph{fly-by} experimental scenario where one of the two qubits streams past the other one with a given velocity $\bm{v}$. While not the dominant model in current demonstrations of digital quantum computation with neutral atoms, this setup is becoming increasingly relevant in recent proposals for \emph{streaming} architectures where on-demand qubit transport is replaced with global streams of constant-velocity atoms enabling processor connectivity~\cite{Dudinets2026AlltoallConnectivityRydbergatombased,lib2026velocityenabledquantumcomputingneutral}.

The relative velocity of the two atoms gives rise to a differential Doppler shift $\delta_- = \bm{k_{\mathrm{eff}}} \cdot \bm{v}$ where $\bm{k_{\mathrm{eff}}}$ is the effective wavevector. For a two-photon transition, the two lasers are typically counter-propagating in order to minimize $|\bm{k_{\mathrm{eff}}}|$. Nevertheless, assuming two-photon excitation via $6P_{3/2}$ in rubidium-87 (wavelengths 420 and 1015 nm) with a velocity $|\bm{v}| = 1 \ \mathrm{m \, s^{-1}}$, the resulting Doppler shift is $\sim 2\pi \times1.4~ \mathrm{MHz}$, which cannot be neglected with respect to typical Rydberg Rabi frequencies. Aligning the Rydberg beams orthogonal to $\bm{v}$ to suppress the Doppler shift would represent an important architectural constraint.

Proposals have been made to address the asymmetric detuning via asymmetric atom addressing, two lasers with opposite wavevectors~\cite{Shi2020SuppressingMotionalDephasing,cole2026asymmetricfastrydberggate}, or Doppler-robust pulses based on motional control. However, because atoms are transported at a deterministic velocity, the Doppler shift of each atom can be calibrated with high accuracy. Further, Doppler shift errors due to finite atomic temperature are identical to those in a conventional Rydberg gate via a reference frame change. Therefore, it is sufficient to tailor a gate to a specific detuning $\delta_-$: in this context, we find that there exists a simple, shorter, $\ket{1}\leftrightarrow \ket{r}$ protocol that implements a $CZ_{\theta}$ entangling gate.

\subsection{Gate requirements and results}

The level diagram is written in the common-mode frame where the static atom (in the lab frame) has detuning $-\delta_-/2$, and the moving atom $\delta_-/2$. The resulting fly-by Rydberg Hamiltonian is still block-diagonal $H'^{R} = H_{00}^R \oplus H_{01}'^R \oplus H_{10}'^R \oplus H_{11}'^R$ with blocks:
\begin{equation}
H_{01}'^R =
\begin{pmatrix}
0 & \frac{\Omega}{2}\\
\frac{\Omega^*}{2} & \delta_-/2
\end{pmatrix}
\text{ in basis } \{\ket{01},\ket{0r}\};
\end{equation}
\begin{equation}
H_{10}'^R =
\begin{pmatrix}
0 & \frac{\Omega}{2}\\
\frac{\Omega^*}{2} & -\delta_-/2
\end{pmatrix}
\text{ in basis } \{\ket{10},\ket{r0}\};
\end{equation}
\begin{equation}
H_{11}'^R =
\begin{pmatrix}
0 & \frac{\Omega}{\sqrt{2}} & 0\\
\frac{\Omega^*}{\sqrt{2}} & 0 & \delta_-/2\\
0 & \delta_-/2 & 0
\end{pmatrix}
\text{ in basis } \{\ket{11},\ket{W}, \ket{D}\};
\end{equation}
as summarized in Fig.~\ref{fig:fly-by}(a). The differential detuning therefore results in a coupling between the Rydberg bright state $\ket{W}$ and dark state $\ket{D}$, as well as an extra diagonal $\delta_-/2$ term in the single-excitation blocks.

When considering the $\delta_-$ terms as perturbations on top of the original Rydberg protocol's requirements, two additional constraints emerge on the unitary $U$ (Appendix~\ref{app:flyby}):
\begin{equation}
\int_0^T{\langle W|U(t)|11\rangle dt} = 0;\label{eq:flyby-block1}
\end{equation}
\begin{equation}
\int_0^T{\langle 0r|U(T) U(t)^{\dagger}|0r\rangle \langle 0r|U(t) |01\rangle dt} = 0.\label{eq:flyby-block2}
\end{equation}

Because these constraints do not involve $\delta_-$, the required gate duration jumps from $\sim 7.6/|\Omega|$ for $\delta_- = 0$ (regular time-optimal gate) to $\sim 12.7/|\Omega|$ for any $0 < |\delta_-| \ll |\Omega|$. More generally, the required gate duration for $\delta_- \in [0,|\Omega|]$ is shown in Fig.~\ref{fig:fly-by}(b), along with the gate profile for $\delta_- = |\Omega|/2$.

The phase profile found via numerical optimization is odd up to a constant offset, i.e.,
\begin{equation}
\phi(t) + \phi(T-t) \approx \mathrm{const},
\end{equation}
which automatically satisfies two of the four real-valued constraints of Eq.~\ref{eq:flyby-block1} and Eq.~\ref{eq:flyby-block2}.

Note that the no-go theorem of Sec.~\ref{sec:physical-setup} still applies to the fly-by gate for common-mode detuning errors, since $\ket{00}$ remains dark. Finally, in Appendix~\ref{app:more-gates}, we present composite robust and pseudo-robust versions of the fly-by gates for $\delta_- \in [0,|\Omega|]$, respectively using $T|\Omega| \in [28.7,36.6]$ and $T|\Omega| \in [17.0,22.3]$.

\section{Conclusion}\label{sec:conclusion}

Our work addresses a central obstacle to the scalability of parallel Rydberg entangling gates. We have shown that the detuning sensitivity of Rydberg entangling gates, which no conventional protocol can suppress, can indeed be removed by involving both qubit states in the gate. The mechanism, which can be realized either via the composite (or co-driven) or pseudo-robust protocols delineated in this work, rests on a single condition -- the Rydberg-dwell balance $N_{11} = N_{01} + N_{10}$ -- that converts detuning errors into $ZI+IZ$ errors.

While these protocols come with trade-offs, the asymptotic dominance of light-shift errors in two-photon Rydberg gate error budgets indicates that they could provide a several-fold improvement over the conventional time-optimal gate -- particularly as laser power continues increasing, at which point even sub-percent intensity inhomogeneity may become a hurdle to reaching ultra-high fidelities with conventional Rydberg protocols.

We also introduce the notion of pseudo-robustness, where an error is converted into a benign, correctable form rather than suppressed -- in the same spirit as erasure conversion~\cite{Wu2022ErasureConversionRydberg} -- which can significantly lower the required gate time for error reduction when outright robustness is too costly or unachievable. Our work also addresses the problem of realizing fly-by gates in emerging streaming computing architectures -- without requiring the machinery of Doppler-robustness.

Finally, we provide an extensive set of characterizations (notably in the Appendices) that include static error sensitivities, fidelity response to AC laser noise~\cite{tsaiBenchmarkingFidelityResponse2025}, and the protocols' Hessian waveforms that facilitate experimental \textit{in-situ} fidelity optimization~\cite{liu2026highfidelityneutralatomgates}.

\textit{Note added.} During completion of this manuscript, we became aware of a related work~\cite{Covey2026companion}.

\vspace{-0.3cm}
\begin{acknowledgments}
We acknowledge insightful discussions with, and feedback from, Kangheun Kim, Kon Leung, Hannah Manetsch, Nadine Meister, Xiangkai Sun, Richard Tsai, Dolev Bluvstein, and Simon Evered. We thank Jacob Covey, Zhubing Jia, and Yichao Yu for informing us of their ongoing work~\cite{Covey2026companion}. We acknowledge support from the Heising-Simons Foundation (20244852); the NSF QLCI program (2016245); the Institute for Quantum Information and Matter, an NSF Physics Frontiers Center, (NSF Grant No PHY-1733907); the AFOSR (FA9550-23-1-0625); the Army Research Office TINA QC program (W911NF2410388) and Oratomic. Support of this work is also acknowledged from the U.S. Department of Energy, Office of Science, National Quantum Information Science Research Centers, Quantum Systems Accelerator. E.B. acknowledges support from the Eddleman Graduate Fellowship.

\textit{Competing Interests:} M.E. is a co-founder, shareholder, and advisor of Oratomic.
\end{acknowledgments}

\FloatBarrier
\newpage
\appendix
\setcounter{figure}{0}
\renewcommand{\thefigure}{A\arabic{figure}}
\renewcommand{\theHfigure}{appendix.\arabic{figure}}

\section{Detailed scaling of error channels}
\label{app:error_scalings}

Here we provide a discussion of the scaling of the dominant
error channels in a Rydberg gate with a two-photon transition. The purpose of this section is not to provide the exact numerical prefactors, which depend on the pulse shape and gate protocol, but rather to isolate the dependence on the
experimentally relevant parameters, and illustrate trade-offs in parameter choices associated with detuning-induced errors.

\subsection{Reparameterization of the two-photon coupling}
We consider the ladder system $\ket{1}\leftrightarrow\ket{e}\leftrightarrow\ket{r}$, with single-photon Rabi frequencies $\Omega_{1e}$ and $\Omega_{er}$ and single-photon detuning $\Delta$. In the far-detuned regime, $|\Delta|\gg |\Omega_{1e}|,|\Omega_{er}|$, adiabatic elimination of $\ket{e}$ gives
\begin{equation}
    \Omega \equiv \Omega_{\mathrm{eff}} \simeq \frac{\Omega_{1e}\Omega_{er}}{2\Delta}.
    \label{eq:Omega_eff_app}
\end{equation}
Defining the drive imbalance $\beta\equiv|\Omega_{1e}/\Omega_{er}|$, we obtain
\begin{align}
    |\Omega_{1e}|^2 &\simeq 2|\Omega\Delta|\,\beta,\\
    |\Omega_{er}|^2 &\simeq 2|\Omega\Delta|\,\beta^{-1}.
\end{align}
We therefore parameterize the gate by $\{\Omega,\beta,\Delta,n\}$, where $n$ is the principal quantum number of the Rydberg state. For a fixed gate protocol, $T\propto|\Omega|^{-1}$.

Since $I_{ij}\propto |\Omega_{ij}|^2/|\langle j|d|i\rangle|^2$, the required lower-leg intensity scales as
\begin{equation}
    I_{1e} \propto |\Omega||\Delta|\beta,
    \label{eq:I_1e_scaling}
\end{equation}
where the relevant dipole matrix element is approximately independent of $n$. For the upper leg, $|\langle r|d|e\rangle|\propto n^{-3/2}$ ~\cite{gallagherRydbergAtoms1994,sibalicARCOpensourceLibrary2017}, giving
\begin{equation}
    I_{er} \propto |\Omega||\Delta|\frac{n^3}{\beta}.
    \label{eq:I_er_scaling}
\end{equation}

\subsection{Dissipative errors}
\paragraph{Intermediate-state scattering.}

In the far-detuned regime, the intermediate state is only weakly populated
and can be adiabatically eliminated. Its population is generated coherently
through the couplings from both \(\ket{1}\) and \(\ket{r}\), and therefore
contains not only terms proportional to
\(|\Omega_{1e}|^2\) and \(|\Omega_{er}|^2\), but also an interference term
between the two excitation pathways. After averaging over the gate, this gives $\overline{P_e}\propto (c_1|\Omega_{1e}|^2+c_r|\Omega_{er}|^2+c_{\mathrm{cross}}|\Omega_{1e}\Omega_{er}|)/\Delta^2$, where the coefficients are protocol-dependent and of order unity, and $c_{\mathrm{cross}}$ parameterizes the destructive or constructive interference between the two pathways.

Using Eq.~\ref{eq:I_1e_scaling},~\ref{eq:I_er_scaling} together with \(T\propto|\Omega|^{-1}\), the corresponding scattering error, with \(\Gamma_e\) the decay rate of \(\ket{e}\), scales as
\begin{equation}
    \varepsilon_e \sim \Gamma_e T\overline{P_e} \propto \frac{\Gamma_e}{|\Delta|} \left( c_1\beta + c_r\beta^{-1} + c_{\mathrm{cross}} \right).
    \label{eq:eps_e_scaling}
\end{equation}
Thus, while the detailed dependence on the drive imbalance \(\beta\) is protocol dependent, the leading scaling with the intermediate-state detuning remains
\begin{equation}
    \varepsilon_e \propto \frac{\Gamma_e}{|\Delta|}.
\end{equation}

\paragraph{Rydberg-state decay.}
The Rydberg-state decay error scales as
\begin{equation}
    \varepsilon_r \sim \Gamma_r T \overline{P_r} \propto \frac{1}{n^3|\Omega|},
    \label{eq:eps_r_scaling}
\end{equation}
where \(\overline{P_r}\) is an order-unity, protocol-dependent time-averaged Rydberg population. Here, we use the radiative-lifetime scaling \(\Gamma_r \propto n^{-3}\)~\cite{beterovQuasiclassicalCalculationsBlackbodyradiationinduced2009,gallagherRydbergAtoms1994}.
At finite temperature, blackbody-radiation-induced transitions weaken this scaling; at room temperature, the effective dependence is typically between the radiative \(n^{-3}\) scaling and an approximate \(n^{-2}\) scaling.

\subsection{Quasi-static detuning noise}
For quasi-static detuning \(\delta\), the deviation of the state from the ideal evolution is linear in \(\delta T\), and the resulting infidelity is therefore quadratic in \(\delta T\), so the gate infidelity scales to leading order as
\begin{equation}
    \varepsilon_f
    \propto
    \sigma_\delta^2 T^2
    \propto
    \frac{\sigma_\delta^2}{|\Omega|^2},
    \label{eq:eps_freq_scaling}
\end{equation}
where \(\sigma_\delta\) is the RMS detuning fluctuation.

\subsection{Quasi-static intensity noise}
Intensity fluctuations perturb several terms in the effective two-photon Hamiltonian. For each laser, a fractional intensity fluctuation modulates, to the leading order, (i) the effective two-photon Rabi frequency, (ii) the differential light shift between the qubit states, and (iii) the differential light shift between \(\ket{1}\) and \(\ket{r}\), i.e., the effective Rydberg detuning.
We denote by \(\Delta_{\mathrm{LS},ab}^{(ij)}\) the differential light shift between states \(\ket{a}\) and \(\ket{b}\) induced by the laser driving \(i\leftrightarrow j\). 

The differential light shifts scale as $\Delta_{\mathrm{LS},ab}^{(ij)} \propto \alpha_{ab}^{(ij)} I_{ij}$, where $\alpha_{ab}^{(ij)}$ is the corresponding differential polarizability. For quasi-static intensity fluctuations, a perturbation of characteristic strength \(A\) acting over the gate duration contributes at order \((AT)^2\) to the infidelity. When several terms are present, their effects are coherent and cross terms can contribute; the scaling arguments below therefore refer to the combined perturbation.

\paragraph{Lower-leg laser intensity noise.}

The lower-leg laser (i.e. laser $1 \leftrightarrow e$) intensity fluctuation modifies the two-photon Rabi frequency, the qubit differential light shift \(\Delta_{\mathrm{LS},01}^{(1e)}\), and the Rydberg-resonance shift \(\Delta_{\mathrm{LS},1r}^{(1e)}\). The Rabi-frequency contribution scales as \(|\Omega|\). The relevant lower-leg differential polarizabilities are dominated by the coupling from $\ket{0}, \ket{1}$ to the intermediate state $\ket{e}$.
In the experimentally relevant regime where \(|\Delta|\) is comparable to the hyperfine splitting \(\omega_{\mathrm{hf}}\), the relevant differential polarizabilities scale approximately as \(1/|\Delta|\), up to an order-unity dependence on \(\Delta/\omega_{\mathrm{hf}}\).
Using Eq.~\ref{eq:I_1e_scaling}, the corresponding light shifts therefore scale as
\begin{equation}
    |\Delta_{\mathrm{LS},01}^{(1e)}|, |\Delta_{\mathrm{LS},1r}^{(1e)}| \sim |\Omega|\beta,
\end{equation}
up to order-unity dependence on \(\Delta\).

The Rabi-frequency and light-shift contributions are generally comparable, so no single term needs to dominate. Since the former is approximately independent of \(\beta\), while the latter scale approximately as \(\beta\), their combined contribution to the infidelity can in general be written as
\begin{equation}
    \varepsilon_{I,1e} \sim \sigma_{I,1e}^2 (C_0+C_1\beta+C_2\beta^2) ,
    \label{eq:eps_I_lower_scaling}
\end{equation}
where \(C_0\) originates from the Rabi-frequency fluctuation, \(C_1\) from its coherent cross terms with the light-shift perturbations, and \(C_2\) from the light-shift perturbations themselves, including their mutual cross term. $\sigma_{I,1e}$ is the RMS of the relative intensity noise on laser $1 \leftrightarrow e$. The coefficients are protocol dependent, and \(C_1\) can have either sign; consequently, the dependence on \(\beta\) need not be monotonic. Nevertheless, for \(\beta=O(1)\), all terms remain of the same order, and since \(T\propto|\Omega|^{-1}\), the lower-leg intensity-noise error has no leading dependence on \(|\Omega|\) and only
weak dependence on \(|\Delta|\).

\paragraph{Upper-leg laser intensity noise.}

For the upper-leg laser (i.e. laser $e \leftrightarrow r$), the qubit differential light shift \(\Delta_{\mathrm{LS},01}^{(er)}\) is negligible compared to the other terms. The relevant contributions are the fluctuation of the two-photon Rabi frequency and the Rydberg-resonance shift \(\Delta_{\mathrm{LS},1r}^{(er)}\). The former again scales as \(|\Omega|\), and hence gives an approximately parameter-independent contribution after multiplication by the gate duration.

The Rydberg-resonance shift scales as $\Delta_{\mathrm{LS},1r}^{(er)} \propto \alpha_{1r}^{(er)} I_{er}$. In the relevant large-detuning regime, \(\alpha_{1r}^{(er)}\) is mostly dominated by the Rydberg ponderomotive polarizability and the ground-state polarizability, and varies only weakly with \(n\) and \(\Delta\). Using the upper-leg intensity scaling derived above (Eq.~\ref{eq:I_er_scaling}), we obtain
\begin{equation}
    |\Delta_{\mathrm{LS},1r}^{(er)}| \propto |\Omega||\Delta|\frac{n^3}{\beta}.
\end{equation}

For sufficiently large detuning, this resonance-shift contribution becomes larger than the Rabi-frequency contribution and determines the leading parameter dependence. Using \(T\propto |\Omega|^{-1}\), the resulting asymptotic scaling is
\begin{equation}
    \varepsilon_{I,er} \propto \sigma_{I,er}^2 \frac{\Delta^2 n^6}{\beta^2} ,
    \label{eq:eps_I_upper_scaling}
\end{equation}
where $\sigma_{I,er}$ is the RMS of the relative intensity noise on laser $e \leftrightarrow r$. At smaller detuning, the Rabi-frequency contribution can become comparable or dominant, giving instead an approximately parameter-independent contribution; this does not alter the large-\(|\Delta|\) scaling above.

\subsection{Stark correlation and intensity-noise sensitivity}
\label{app:stark_correlation}

The intensity-noise scalings derived above can be expressed more compactly
using the Stark correlation introduced in the main text (also see Ref.~\cite{janduraOptimizingRydbergGates2023}). To distinguish the
different differential light shifts relevant here, we define
\begin{equation}
    \zeta_{ab}^{(ij)} \equiv \frac{2\Delta_{\mathrm{LS},ab}^{(ij)}}{|\Omega|}.
    \label{eq:zeta_general}
\end{equation}
The Stark correlation used in the main text is therefore
\begin{equation}
    \zeta_{ij} \equiv \zeta_{1r}^{(ij)}.
\end{equation}

The differential light shift is proportional to the corresponding differential polarizability and laser intensity,
\begin{equation}
    \Delta_{\mathrm{LS},ab}^{(ij)} \propto \alpha_{ab}^{(ij)} I_{ij},
\end{equation}
such that
\begin{equation}
    \zeta_{ab}^{(ij)} \propto \frac{\alpha_{ab}^{(ij)} I_{ij}}{|\Omega|}.
    \label{eq:zeta_alpha}
\end{equation}
Thus, \(|\zeta|\) directly measures the strength of an intensity-induced differential light shift relative to the two-photon coupling. Since \(T\propto|\Omega|^{-1}\), the light-shift contribution to the quasi-static intensity-noise error scales as
\begin{equation}
    \varepsilon_I^{(\mathrm{LS})} \propto \sigma_I^2 \zeta^2,
    \label{eq:eps_zeta}
\end{equation}
up to protocol-dependent response factors, where $\sigma_I$ denotes the RMS fractional intensity fluctuation.

For the lower-leg laser, the relevant differential polarizabilities are dominated by the near-resonant coupling to the intermediate state and scale approximately as \(1/|\Delta|\) over the regime of interest. Together with \(I_{1e}\propto|\Omega||\Delta|\beta\), this gives
\begin{equation}
    \zeta_{01}^{(1e)}, \zeta_{1r}^{(1e)} \sim O(\beta),
    \label{eq:zeta_lower}
\end{equation}
up to order-unity dependence on \(\Delta\). This makes explicit why the lower-leg light-shift sensitivity does not acquire a strong dependence on either \(|\Omega|\) or \(|\Delta|\). As discussed above, however, the Rabi-frequency and light-shift perturbations are comparable for the lower-leg laser, so \(\zeta_{1r}^{(1e)}\) alone does not determine the total lower-leg intensity-noise error.

For the upper-leg laser, the qubit differential light shift is negligible, and at sufficiently large detuning the intensity-noise sensitivity is dominated by the Rydberg-resonance shift \(\Delta_{\mathrm{LS},1r}^{(er)}\). In this regime, the corresponding
differential polarizability is dominated by the Rydberg ponderomotive and ground-state polarizabilities and depends only weakly on \(n\) and \(\Delta\). Using Eq.~\ref{eq:I_er_scaling}, we obtain
\begin{equation}
    \zeta_{er} \equiv \zeta_{1r}^{(er)}, \qquad |\zeta_{er}| \propto \frac{n^3|\Delta|}{\beta}.
    \label{eq:zeta_upper}
\end{equation}
Consequently, in the regime where this light-shift contribution dominates, \(\varepsilon_{I,er}\propto\zeta_{er}^2\), recovering Eq.~\ref{eq:eps_I_upper_scaling}.

\begin{figure*}
\centering
\includegraphics[width=\textwidth]{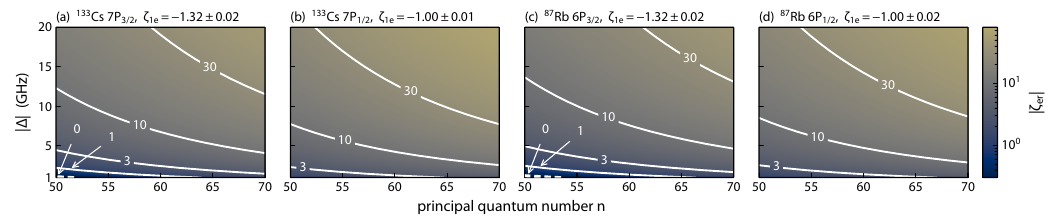}
    \caption{\label{fig:zeta-er}\textbf{Stark-correlation maps for alkali two-photon Rydberg excitation.}
    Stark-correlation maps for balanced one-photon Rabi frequencies, $\beta \equiv |\Omega_{1e}/\Omega_{er}| = 1$, for excitation to $nS_{1/2}$ Rydberg states in four representative alkali implementations: (a) $^{133}\mathrm{Cs}$ via $7P_{3/2}$, (b) $^{133}\mathrm{Cs}$ via $7P_{1/2}$, (c) $^{87}\mathrm{Rb}$ via $6P_{3/2}$, and (d) $^{87}\mathrm{Rb}$ via $6P_{1/2}$. In each panel, the background shading shows the magnitude of the upper-leg Stark correlation, $|\zeta_{er}|$, as a function of the Rydberg principal quantum number $n$ and the magnitude of the intermediate-state detuning $|\Delta|$. Here $|\Delta|$ denotes the magnitude of a negative detuning, $\Delta<0$, corresponding to a red-detuned lower-leg laser and a blue-detuned upper-leg laser. White solid contours indicate representative nonzero values of $|\zeta_{er}|$, while the dotted contour marks $\zeta_{er}=0$, where the upper-leg differential light shift between $|1\rangle$ and $|r\rangle$ vanishes. The lower-leg Stark correlation $\zeta_{1e}$ varies by only a few percent over each panel, with its representative value and variation quoted above each panel. For other drive imbalances, the correlations rescale as $\zeta_{er} \rightarrow \zeta_{er}/\beta$ and $\zeta_{1e} \rightarrow \beta\,\zeta_{1e}$. These maps provide a direct conversion between the dimensionless Stark-correlation parameter used in the main text and experimentally relevant choices of atomic species, intermediate state, Rydberg state, and single-photon detuning.}
\end{figure*}

The Stark correlation therefore provides a compact measure of the gate sensitivity to intensity fluctuations arising from differential light shifts. In particular, the upper-leg intensity-noise sensitivity is reduced by choosing smaller \(n\) and smaller \(\Delta\), in direct competition with dissipative errors, which favor larger \(n\) and larger \(\Delta\). Fig.~\ref{fig:zeta-er} quantifies this trade-off for the standard two-photon schemes in cesium and rubidium; in particular, the parameters of Ref.~\cite{evered2026highfidelityentanglinggatesnonlocal} yield \(|\zeta_{er}| \approx 6\).

\subsection{Experimental implications and practical trade-offs}
\label{app:experimental_implications}

Taken together, several standard strategies for suppressing dissipative errors -- notably increasing $n$ or $|\Delta|$ -- also tend to increase the magnitude of the upper-leg Stark correlation $|\zeta_{er}|$. The large-$|\zeta|$ regime considered in this work is therefore not an artificial corner of parameter space, but emerges naturally when optimizing conventional two-photon Rydberg gates.

At the same time, no single asymptotic scaling determines the optimal operating point: different error mechanisms favor competing regions of parameter space, and several constraints are architectural rather than atomic. A quantitative choice of $\{n,\Delta,\Omega,\beta\}$ therefore requires a more complete treatment of the relevant error mechanisms together with a joint optimization over the experimentally accessible parameters.

One example is finite Rydberg blockade. At fixed interatomic separation, the van der Waals interaction scales as $C_6\propto n^{11}$, strongly favoring larger $n$, as the corresponding finite-blockade error scales as $\varepsilon_B\sim(|\Omega|/B)^2$ (although amplitude shaping can further suppress the doubly-excited population). Other effects favor the opposite direction in $n$: the Rydberg electric polarizability grows approximately as $n^7$, and the increasing density of nearby Rydberg levels can enhance off-resonant coupling to non-target states. The choice of $n$ can also constrains how densely different gate pairs can be operated in parallel, an architectural consideration. Other detuning errors, including those arising from stray electric fields, Doppler shifts, and laser frequency noise, can also introduce additional trade-offs beyond the light-shift mechanism considered above.

\section{Numerical methods}\label{app:numerics}

\subsection{Optimization formalism}

The gate protocols presented in this work are optimized with task-specific C++ programs, and all reported quantities are then re-evaluated with an independent, general Julia implementation. All quantities are expressed in units of the Rydberg Rabi frequency, $|\Omega| = 1$.

The task-specific optimization algorithms use the Bell state fidelity for convenience:
\begin{equation}
    F_B(U,U_t)=\left|  \frac{1}{D} \mathrm{tr}(P U_t^\dagger U)\right|^2,
\end{equation}
although all the fidelities and sensitivities reported in this work are subspace-averaged (Eq.~\ref{eq:fidelity}).

We define \emph{main} control parameters $\bm{u}(t_l)$ for each time step $t_l$ and \emph{additional} control parameters $\bm{u}'$ (e.g., the single-qubit phase $\theta$, the gauge angles of Eq.~\ref{eq:local-target} and $\kappa$), as well as error channels $a$. The Hamiltonian for time step $t_l = T(l-1)/L$ ($l\in \llbracket 1,L \rrbracket$, with $L=200$ per pulse, i.e., per half-gate for the composite gate) is given by
\begin{equation}
H_l = H_0(t_l,\bm{u}(t_l),\bm{u}') + \sum_a{\varepsilon_a V_a(t_l,\bm{u}(t_l),\bm{u}')},
\end{equation}
with errorless Hamiltonian $H_0$ and error operators $\{V_a\}$ ($V_\delta = P_r$ and $V_\epsilon = \hat{D}$ for the detuning and amplitude errors). The conventionally driven protocols -- time-optimal, pseudo-robust, and the composite half-gate -- are optimized in the 5-state symmetric basis $\{\ket{00}, \ket{01}, \ket{11}, \ket{0r}, \ket{W}\}$, where the symmetric combination of $\ket{01}$ and $\ket{10}$ appears once with weight 2 in the fidelity. This basis is exact for the composite gate, because the half-gate returns all population to the computational subspace before each $X^{\otimes 2}$, so that $\ket{D}$ is never populated. The co-driven gate is optimized in the 8-state basis $\{\ket{0},\ket{1},\ket{r}\}^{\otimes 2}\setminus\{\ket{rr}\}$, and the fly-by gate in the blocks of Sec.~\ref{sec:flyby}.

\subsection{Propagators}

For the conventionally driven protocols, the GRAPE propagators can be written exactly thanks to the block-diagonal form of the Hamiltonian: with $\omega = \sqrt{\Omega_X^2 + \delta^2/4}$ and $\Omega_X = \chi_q(1+\epsilon)|\Omega|/2$,
\begin{equation}
e^{-iH_l dt} = e^{-i\delta dt/2}
\begin{pmatrix}
c + i\frac{\delta}{2\omega}s & -i\frac{\Omega_X}{\omega}s\, e^{i\phi_l} \\
-i\frac{\Omega_X}{\omega}s\, e^{-i\phi_l} & c - i\frac{\delta}{2\omega}s
\end{pmatrix}
\end{equation}
in the basis $\{\ket{q}, \ket{r(q)}\}$, with $c = \cos\omega dt$ and $s = \sin\omega dt$. For the co-driven gate, the propagator is factorized exactly as $e^{-iH dt} = G^\dagger e^{-iH_{\mathrm{real}} dt} G$, where $G = \mathrm{diag}(e^{i\nu_k})$ collects the drive phases ($\nu_0 = 0$, $\nu_1 = \psi$, $\nu_r = \psi + \phi$, additive over the two atoms) and $H_{\mathrm{real}}$ is the real symmetric Hamiltonian with both phases gauged away; $e^{-iH_{\mathrm{real}} dt}$ is evaluated by a Taylor series of order 24. The qubit-drive amplitude is parameterized as $|\Omega_1| = \Omega_{1,\max}\sin^2 u$, so that both bounds are attainable and the gradient never saturates.

\subsection{Cost functions}

All cost functions have the form $C = 1 - F_B(U, U_t) + \lambda E + R$, where $E$ penalizes the first-order response to the error and $R$ is a small roughness regularization. For the composite half-gate, the full cost reads
\begin{align}
C_{\mathrm{comp}} &= 1- F_B(V,R_{\theta}^{\otimes 2} C_{\pi/2})\nonumber\\
&\quad +\lambda \biggl(\sum_q{\sum_{k\neq q} |\langle k|\psi_q^{(1)}\rangle|^2}\nonumber\\
&\quad +\left|2e^{-i\theta} \langle 01|\psi_{01}^{(1)}\rangle-e^{-i(2\theta + \pi/2)} \langle 11| \psi_{11}^{(1)}\rangle \right|^2 \biggr),\label{eq:half-gate-req}
\end{align}
where $\ket{\psi_q} = V\ket{q}$ and $\ket{\psi_{q}^{(1)}} = \partial |\psi_q\rangle/\partial\delta$. The weight $\lambda$ (here $10^{-4}$) is calibrated to ensure that the resulting unitaries always achieve $F>1-10^{-5}$. The pseudo-robust gate uses the same penalties with a $CZ$ target in place of $C_{\pi/2}$; for amplitude-and-detuning robustness, the same two terms are added for the amplitude response. For the co-driven gate, the cost is
\begin{align}
C_{\mathrm{co}} &= 1-F_B(U_t,U)\nonumber\\
&\quad + \lambda \sum_q{\sum_k{|\langle k|\psi_q^{(1)}\rangle + i\kappa \langle k|\psi_q^{(0)}\rangle|^2}},\label{eq:codriven-cost}
\end{align}
where $\ket{\psi_q^{(0)}} = U\ket{q}$, the sum over $k$ runs over all 8 states, and the global phase gauge $\kappa$ is optimized jointly; the weight $\lambda \in [10^{-4},10^{-2}]$ is tuned to achieve different trade-offs of robustness and fidelity. The regularization is $R = w_1\sum_l |e^{i\phi_{l+1}} - e^{i\phi_l}|^2 + w_2\sum_l |e^{i\phi_{l+1}} - 2e^{i\phi_l} + e^{i\phi_{l-1}}|^2$, with $w_1 = w_2 = 10^{-6}$ for the conventionally driven protocols, and, for the co-driven gate, $w_1 = 3\times10^{-8}(L/T)^2$ on both phases together with an analogous penalty $3\times10^{-7}(L/T)^4$ on the second differences of the qubit-drive amplitude.

\subsection{Gradients, optimizer and seeds}

Gradients are obtained by automatic differentiation (Enzyme~\cite{Moses2020InsteadRewritingForeignCode}) through the closed-form step propagators and the cost is minimized with L-BFGS~\cite{Liu1989LimitedMemoryBFGS}. Depending on the protocol, we use multiple random seeds, L-BFGS restarts, and continuation (seeding a gate optimized for duration $T$ with the one found at neighboring $T'$) to improve convergence. For the co-driven gate, whose optimization landscape is significantly more rugged than the composite one, 10 seeds with 3 restarts are used at each $T$, together with continuation.

\subsection{Onset criterion}

\begin{figure}
\centering
    \includegraphics[width=86mm]{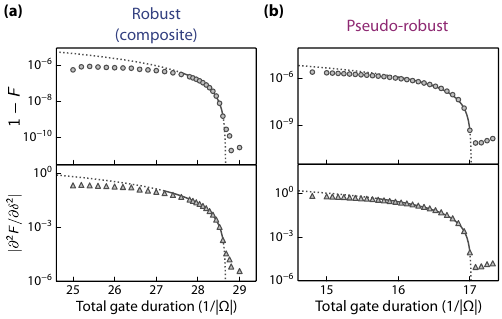}
    \caption{\label{fig:sm-convergence} Convergence of (a) the detuning-robust composite gate and (b) the detuning pseudo-robust gate: infidelity and detuning sensitivity versus total gate duration, with the quadratic fits (solid lines) yielding the onsets $T_0 \approx 28.67/|\Omega|$ and $17.02/|\Omega|$.}
\end{figure}

Near the onset $T_0$ of a robust or pseudo-robust protocol, the infidelity and the sensitivities vanish quadratically, $y(T) = y_0 (T - T_0)^2$ (Fig.~\ref{fig:sm-convergence}). The durations quoted in Table~\ref{tab:comparison} are the scan grid points just above onsets found via the quadratic fit.

\section{Additional gate protocols}\label{app:more-gates}

\emph{Amplitude- and detuning-robust composite gate.} Adding the amplitude error to the cost of Eq.~\ref{eq:half-gate-req}, i.e., requiring the absence of leakage and the Rydberg-dwell balance for the first-order response to $\epsilon$ as well as to $\delta$ (with $D_q$ in place of $N_q$, Appendix~\ref{app:corrected}), yields a composite gate robust to both errors. Quadratic fits of the convergence give onsets of $T_0 = 43.3/|\Omega|$ for the detuning sensitivity and $46.4/|\Omega|$ for the amplitude sensitivity; the gate of Table~\ref{tab:comparison} ($T = 46.48/|\Omega|$) lies above both onsets, with $S_\epsilon$ and $|\Omega|^2 S_\delta$ below $10^{-3}$.

\emph{Amplitude- and detuning-pseudo-robust gate.} The same extension of the pseudo-robust cost yields a gate that is pseudo-robust to both errors for $T \gtrsim 24.27/|\Omega|$ (Table~\ref{tab:comparison}). As noted in Sec.~\ref{sec:pseudo}, this gate turns out to be amplitude-robust as well ($S_\epsilon = 3.5\times10^{-3}$), which is not required by the cost function.

\emph{Finite Stark correlation.} Fig.~\ref{fig:stark}(a) shows the onset duration $T^*(\zeta)$ of the pseudo-robust and composite gates optimized for the correlated error operator $\partial H^R/\partial\epsilon$ of Sec.~\ref{sec:physical-setup}. For any finite $\zeta$, both gates are shorter than their detuning-only limit, which they approach as $T^*(\zeta) \simeq T^*(\infty) - a/\zeta$, with $a \approx 9.1/|\Omega|$ for the pseudo-robust gate ($T^*(\infty) = 17.0/|\Omega|$) and $a \approx 19/|\Omega|$ for the composite gate ($T^*(\infty) \approx 28.8/|\Omega|$, consistent with the detuning-only onset within the fit uncertainty).

\emph{Fly-by variants.} Pseudo-robust and composite-robust versions of the fly-by gate of Sec.~\ref{sec:flyby} -- robust to a common-mode detuning error at a fixed differential detuning $\delta_-$ -- exist for $\delta_- \in [0, |\Omega|]$, with durations $T|\Omega| \in [17.0, 22.3]$ and $[28.7, 36.6]$, respectively.
\section{Equivalence of co-driven, two-tone and EOM schemes}\label{app:two-tone}

\emph{Two-tone drive.} Consider a single atom driven on both $\ket{0}\leftrightarrow\ket{r}$ and $\ket{1}\leftrightarrow\ket{r}$ with complex Rabi frequencies $\omega_0 = |\omega_0|e^{i\phi_0}$ and $\omega_1 = |\omega_1|e^{i\phi_1}$, i.e., $h = \frac{1}{2}\left(\omega_0\ket{0} + \omega_1\ket{1}\right)\bra{r} + \mathrm{h.c.}$
Defining the bright qubit state $\ket{b} = (\omega_0\ket{0} + \omega_1\ket{1})/\Omega_b$ with $\Omega_b = \sqrt{|\omega_0|^2 + |\omega_1|^2}$, and dark state $\ket{d}$ in the qubit subspace, the drive reads $h = \frac{\Omega_b}{2}\ket{b}\bra{r} + \mathrm{h.c.}$: only $\ket{b}$ is coupled to $\ket{r}$.
With $\tan\eta = |\omega_0|/|\omega_1|$ and $\Delta\phi = \phi_0 - \phi_1$, $\ket{b} = e^{i\phi_1}\left(\sin\eta\, e^{i\Delta\phi}\ket{0} + \cos\eta\ket{1}\right)$ and $\ket{d} = \cos\eta\ket{0} - \sin\eta\, e^{-i\Delta\phi}\ket{1}$.

Let $W(t)$ be the unitary mapping $\ket{0} \to \ket{d(t)}$, $\ket{1} \to \ket{b(t)}$ and $\ket{r} \to \ket{r}$. In the frame defined by $W$, the Hamiltonian becomes $h' = W^\dagger h W - i W^\dagger \dot{W}$. The first term is a conventional Rydberg drive of amplitude $\Omega_b$ on $\ket{1}\leftrightarrow\ket{r}$. The second term acts within the qubit subspace: its off-diagonal element is
\begin{equation}
-i\langle d|\dot{b}\rangle = -i e^{i(\phi_1 + \Delta\phi)}\left[\dot{\eta} + \frac{i}{2}\Delta\dot{\phi}\sin 2\eta\right],
\end{equation}
and its diagonal elements amount to a $z$-rotation of the qubit, which can be absorbed into the phases $\phi(t)$ and $\psi(t)$. The two-tone scheme is therefore equivalent to a co-driven gate with single-qubit Rabi frequency
\begin{equation}
|\Omega_1| = \sqrt{4\dot{\eta}^2 + \Delta\dot{\phi}^2 \sin^2 2\eta}.
\end{equation}

For a fixed power splitting ($\dot{\eta} = 0$), $|\Omega_1| = |\dot{\phi}_0 - \dot{\phi}_1| \sin 2\eta$: the bound $\Omega_{1,\mathrm{max}}$ on the single-qubit drive becomes a bound on the slew rate of the relative phase of the two tones, $|\dot{\phi}_1 - \dot{\phi}_0|$. The robust composite gate can be realized by driving the first half-gate on $\ket{1}\leftrightarrow\ket{r}$ ($\eta = 0$) and the second on $\ket{0}\leftrightarrow\ket{r}$ ($\eta = \pi/2$). In practice, this requires combining two beams separated by the qubit frequency, typically a few GHz for hyperfine qubits.

\emph{EOM modulation.} When $\ket{0}$ and $\ket{1}$ are the magnetically insensitive states of a hyperfine qubit (e.g., in rubidium-87 or cesium-133), the two tones can be generated by modulating a single $\ket{1}\leftrightarrow\ket{r}$ laser with an electro-optic modulator (EOM) at the qubit splitting frequency.
The carrier and first sideband then have Rabi frequencies $\omega_1 = \Omega J_0(\beta_{\mathrm{mod}})$ and $\omega_0 = \Omega J_1(\beta_{\mathrm{mod}}) e^{i\phi_{\mathrm{mod}}}$, where $\beta_{\mathrm{mod}}$ is the modulation depth, $\phi_{\mathrm{mod}}$ the modulation phase, and $J_n$ denotes the Bessel functions of the first kind. Resonant EOMs reach modulation depths of up to $\pi$ at such modulation frequencies and are compatible with optical powers of tens of watts.

Compared with the co-driven scheme, the power in the other sidebands is lost, $\Omega_b = \Omega\sqrt{J_0^2(\beta_{\mathrm{mod}}) + J_1^2(\beta_{\mathrm{mod}})} < \Omega$: the gate duration and the Rydberg dwell time both increase proportionally to $1/\Omega_b$, and with them the decay cost. The EOM scheme can mimic the two-tone composite structure by realizing the first half-gate at $\beta_{\mathrm{mod}} = 0$ ($J_1(\beta_{\mathrm{mod}}) = 0$) and the second at $\beta_{\mathrm{mod}}' \approx 2.4$~rad ($J_0(\beta_{\mathrm{mod}}') = 0$). Since both half-gates must be executed at the same effective Rabi frequency $|\Omega|\min\left(J_0(0), J_1(\beta_{\mathrm{mod}}')\right) \approx 0.52\,|\Omega|$, the total gate time is $T \approx 55.2/|\Omega|$, 93\% longer than the composite gate. A protocol using carrier and first sideband simultaneously reduces this overhead to about 20--25\%.

\section{Error model of the protocol comparison}\label{app:fig5}

\begin{figure*}
\centering
    \includegraphics[width=172mm]{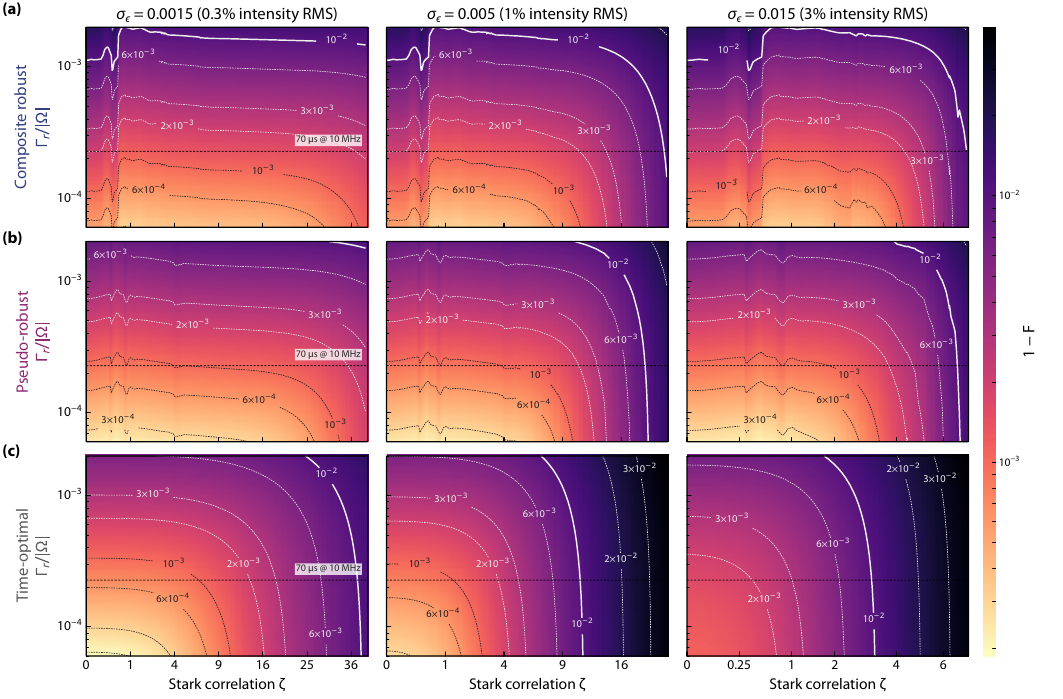}
    \caption{\label{fig:sm-comparison} Absolute infidelity of the Stark-robust composite (top row, $\theta$ frozen), Stark pseudo-robust (middle row, $\theta$ corrected) and time-optimal (bottom row) gates as a function of the Stark correlation $\zeta$ and the normalized Rydberg decay rate $\Gamma_r/|\Omega|$, for $\sigma_\epsilon = 0.0015$, $0.005$ and $0.015$ (columns, corresponding to 0.3\%, 1\% and 3\% intensity RMS). All nine panels share the same logarithmic color scale; the solid contour marks $1 - F = 10^{-2}$ and the dashed horizontal line the reference decay rate $\Gamma_r = 2.3\times10^{-4}\,|\Omega|$.}
\end{figure*}

Fig.~\ref{fig:stark}(b) and Fig.~\ref{fig:sm-comparison} are computed using the exact average gate fidelity, including the quartic term in $\epsilon$ that dominates for the robust and pseudo-robust gates at large detuning $\zeta\epsilon$. For the time-optimal and composite gates the single-qubit phase $\theta$ is fixed to its error-free value, whereas for the pseudo-robust gate $\theta$ is re-optimized at each $\epsilon$, which implements the local phase correction of the pseudo-robust protocol. To estimate the average error for Gaussian-distributed $\epsilon$ with standard deviation $\sigma_{\epsilon}$ we use the 9-point Gauss--Hermite quadrature. Rydberg decay is added at first order,
\begin{equation}\label{eq:full-cost-stark}
1 - F = \left\langle 1 - F(\epsilon) \right\rangle_{\sigma_{\epsilon}} + \Gamma_r \langle N_r \rangle,
\end{equation}
where $\langle N_r \rangle = \frac{1}{4}\sum_q \int_0^T \bra{\psi_q(t)} {P_r} \ket{\psi_q(t)} dt$ is the Rydberg dwell time averaged over the computational basis states.

Because the quartic error is dominant at large $\zeta\epsilon$, but is not directly controlled by the gate optimization, we post-select, among the effective protocols obtained by random-seeding the optimization algorithm, those that yield the highest average fidelity under Eq.~\ref{eq:full-cost-stark}.

\section{Speed limit for Stark-robust conventional gates}\label{app:qsl}

For a conventional protocol at maximal drive, the trajectory starting from $\ket{q}$, $q \in \{01, 10, 11\}$, remains in the block $\{\ket{q}, \ket{r(q)}\}$, $\ket{\psi_q(t)} = c_q(t)\ket{q} + c_{r(q)}(t)\ket{r(q)}$, and the expectation value of the correlated error operator $\partial H^R_q/\partial\epsilon$ of Sec.~\ref{sec:physical-setup} along this trajectory is
\begin{equation}
\bra{\psi_q} \frac{\partial H^R_q}{\partial \epsilon} \ket{\psi_q} = |\Omega| \left[ \zeta p_q + \chi_q \mathrm{Re}\left(e^{i\phi} c_q^* c_{r(q)}\right) \right],
\end{equation}
with $p_q = |c_{r(q)}|^2$. Since $\ket{00}$ is dark, the robustness condition $N_r P = \kappa P$ of Sec.~\ref{sec:physical-setup} (with $N_r$ built from $\partial H^R/\partial\epsilon$ instead of $P_r$) forces $\kappa = 0$, i.e.
\begin{equation}
\zeta N_q + \chi_q \int_0^T \mathrm{Re}\left[e^{i\phi(t)}\, c_q^*(t)\, c_{r(q)}(t)\right] dt = 0.\label{eq:qsl-stark}
\end{equation}

Using $|\mathrm{Re}(e^{i\phi} c_q^* c_{r(q)})| \leq |c_q||c_{r(q)}| = \sqrt{p_q(1-p_q)}$, Eq.~\ref{eq:qsl-stark} implies
\begin{equation}
|\zeta| N_q \leq \chi_q \int_0^T \sqrt{p_q(1-p_q)}\, dt.
\end{equation}
The right-hand side is bounded in two ways. Pointwise, $\sqrt{p_q(1-p_q)} \leq 1/2$, hence $|\zeta| N_q \leq \chi_q T/2$. By the Cauchy--Schwarz inequality, $\int_0^T \sqrt{p_q}\sqrt{1-p_q}\, dt \leq \sqrt{N_q (T - N_q)} \leq \sqrt{N_q T}$, hence $\zeta^2 N_q \leq \chi_q^2 T$. Together these give
\begin{equation}
T \geq N_q \max\left(\frac{2|\zeta|}{\chi_q}, \frac{\zeta^2}{\chi_q^2}\right).\label{eq:qsl-bound-app}
\end{equation}

\section{Corrected fidelity and optimal phase correction}\label{app:corrected}

Let $F(\varepsilon, \vartheta)$ be the fidelity of the gate with error $\varepsilon$ followed by the symmetric phase correction $R_{-\vartheta}^{\otimes 2}$, and let $F_{\varepsilon\varepsilon}$, $F_{\varepsilon\vartheta}$ and $F_{\vartheta\vartheta}$ denote its second derivatives at $(\varepsilon, \vartheta) = (0, 0)$. Since $\partial F/\partial\vartheta = 0$ at this point, the optimal correction for a small error is $\vartheta^*(\varepsilon) = \alpha\varepsilon$ with $\alpha = -F_{\varepsilon\vartheta}/F_{\vartheta\vartheta}$, and substituting it into $F$ gives the corrected curvature
\begin{equation}
\frac{d^2 \tilde{F}}{d\varepsilon^2} = F_{\varepsilon\varepsilon} + 2\alpha F_{\varepsilon\vartheta} + \alpha^2 F_{\vartheta\vartheta} = F_{\varepsilon\varepsilon} - \frac{F_{\varepsilon\vartheta}^2}{F_{\vartheta\vartheta}}.
\end{equation}

For an exact gate without first-order leakage, the error generator is diagonal in the computational basis, $\partial_\varepsilon U = -iU\hat{X}$ with $\hat{X} P = \mathrm{diag}(X_{00}, X_{01}, X_{10}, X_{11})$, where $X_q = N_q$ for a detuning error and $X_q = D_q$ for an amplitude error, so that the phase of $\ket{q}$ drifts as $\theta_q \to \theta_q - X_q\varepsilon$, while the correction adds $(0, -\vartheta, -\vartheta, -2\vartheta)$. To leading order in the combined phase deviations $\varphi_q$ (drift plus correction), $1 - F = \frac{1}{20}\sum_{q<q'}(\varphi_q - \varphi_{q'})^2$, and minimizing over $\vartheta$ yields
\begin{eqnarray}
\vartheta^* &=& \frac{X_{00} - X_{11}}{2}\,\varepsilon, \\
\frac{d^2\tilde{F}}{d\varepsilon^2} &=& -\frac{(X_{11} + X_{00} - 2X_{01})^2}{10}.
\end{eqnarray}
The first expression is the correction quoted in Sec.~\ref{sec:pseudo}, up to the opposite sign convention ($R_{-\vartheta}^{\otimes 2}$ applied here versus $R_{\vartheta}^{\otimes 2}$ there). The second shows that the corrected sensitivity vanishes exactly when $X_{11} + X_{00} = 2X_{01}$. The uncorrected curvature at the same point, $F_{\varepsilon\varepsilon} = -\frac{1}{10}\left[2(X_{01} - X_{00})^2 + (X_{11} - X_{00})^2 + 2(X_{11} - X_{01})^2\right]$, does not vanish: the difference is exactly what the local correction buys.

\section{Fly-by gate: first-order conditions and pulse symmetry}\label{app:flyby}

We write the fly-by Hamiltonian of Sec.~\ref{sec:flyby} as $H'^R = H^R_0 + \frac{\delta_-}{2} K$ with $K = n_r^{(2)} - n_r^{(1)}$, where $n_r^{(i)}$ is the Rydberg projector of atom $i$; the perturbation reads $K = \ket{0r}\bra{0r} - \ket{r0}\bra{r0}$ in the single-excitation blocks and $K = \ket{W}\bra{D} + \ket{D}\bra{W}$ in the $\{\ket{11}, \ket{W}, \ket{D}\}$ block. First-order perturbation theory on $\ket{11}$ gives the dark-state amplitude $\langle D|\psi_{11}(T)\rangle = -\frac{i\delta_-}{2}\int_0^T \bra{D} U_0(T,t)\, K\, U_0(t)\ket{11} dt$. Since $H^R_0\ket{D} = 0$, we have $U_0(T,t)^\dagger\ket{D} = \ket{D}$, so that
\begin{equation}
\langle D|\psi_{11}(T)\rangle = -\frac{i\delta_-}{2} J, \ \text{with } J \equiv \int_0^T \langle W|U_0(t)|11\rangle\, dt:
\end{equation}
the leakage into the dark state is proportional to the time integral of the bright-state \emph{amplitude}, whose phase the control does set. Repeating the argument on $\ket{01}$, the leakage into $\ket{0r}$ is $-\frac{i\delta_-}{2} K'$ with
\begin{equation}
K' \equiv \int_0^T \langle 0r|U_0(T,t)|0r\rangle \langle 0r|U_0(t)|01\rangle\, dt,
\end{equation}
which, unlike $J$, is dressed by the subsequent evolution of $\ket{0r}$; the $\ket{10}$ block gives $-K'$. The diagonal first-order responses of the single-excitation blocks are pure phases, absorbed in $\theta$. First-order robustness to $\delta_-$ is therefore equivalent to $J = K' = 0$, i.e., Eqs.~\ref{eq:flyby-block1} and \ref{eq:flyby-block2}, four real conditions in total.

\section{Cost of the local phase correction}\label{app:correction-cost}

The local correction associated with the pseudo-robust protocol can incur scattering errors whose cost we now compare with the error it removes. For a quasi-static detuning $\delta \sim \mathcal{N}(0, \sigma_\delta^2)$, the pseudo-robust gate channels the error into a symmetric single-qubit phase $\vartheta = (N_{00} - N_{11})\delta/2$, so that the required per-qubit correction has standard deviation $\sigma_\vartheta = (N_{11}/2)\,\sigma_\delta \approx 4.8\,\sigma_\delta$ for the gate of Sec.~\ref{sec:pseudo}. Since a light shift cannot change sign, we consider the following strategy for minimizing the scattering cost: the range $[-s\sigma_\vartheta, s\sigma_\vartheta]$ is corrected with a rotation of one sign, biased by $s\sigma_\vartheta$: the applied angle is at most $2s\sigma_\vartheta$ and $s\sigma_\vartheta$ on average. Tail events ($|\vartheta| > s\sigma_{\vartheta}$) cost at most one full rotation. With $\Gamma_1$ the scattering probability of a $2\pi$ rotation on one qubit, the correction error for both qubits is
\begin{equation}
\varepsilon_c(s) = \frac{\Gamma_1 \sigma_\vartheta}{\pi}\, s\, \mathrm{erf}\!\left(\frac{s}{\sqrt2}\right) + 2\Gamma_1\, \mathrm{erfc}\!\left(\frac{s}{\sqrt2}\right).
\end{equation}
The first term grows linearly with $s$ and the second falls as $e^{-s^2/2}$, so there is an optimum, $s^* \simeq \sqrt{2\ln(1.04\,|\Omega|/\sigma_\delta)} \approx 3$--$4$ in every regime of interest, at which the tails contribute a fraction $1/(s^{*2}+1)$ of the total and
\begin{equation}
\varepsilon_c^* \simeq \frac{\Gamma_1 \sigma_\vartheta}{\pi}\left(s^* + \frac{1}{s^*}\right) = 1.53\,\Gamma_1 \left(s^* + \frac{1}{s^*}\right)\frac{\sigma_\delta}{|\Omega|}.
\end{equation}
The time-optimal gate, in contrast, simply absorbs the error, $\varepsilon_{\mathrm{TO}} = S_\delta \sigma_\delta^2 = 2.83\,\sigma_\delta^2/|\Omega|^2$ (Table~\ref{tab:comparison}). The two scale differently with the noise: the correction costs $O(\sigma_\delta)$, because the angle is linear in the error, while the uncorrected gate costs $O(\sigma_\delta^2)$. The pseudo-robust protocol therefore wins \emph{above} a break-even noise level,
\begin{equation}
\frac{\varepsilon_c^*}{\varepsilon_{\mathrm{TO}}} = 0.54\, \Gamma_1 \left(s^* + \frac{1}{s^*}\right) \frac{|\Omega|}{\sigma_\delta},
\end{equation}
i.e., for $\sigma_\delta > \sigma_\delta^* = 0.54\, \Gamma_1 \left(s^* + 1/s^*\right) |\Omega|$.
If the $Z$ rotation is driven by a laser detuned from an electronic transition of linewidth $\Gamma_{\mathrm{ls}}$, a $2\pi$ qubit phase requires the beam to act for a time set by the hyperfine splitting $\omega_{\mathrm{hf}}$, so that $\Gamma_1 = 2\pi\,\Gamma_{\mathrm{ls}}/\omega_{\mathrm{hf}}$ and $\sigma_\delta^* = 3.4\,(\Gamma_{\mathrm{ls}}/\omega_{\mathrm{hf}})(s^* + 1/s^*)|\Omega|$. For typical schemes in rubidium-87 or cesium-133 (excitation to $6P$ or $7P$), $\Gamma_{\mathrm{ls}}/\omega_{\mathrm{hf}} \sim 2\times10^{-4}$, giving $s^* \approx 3.5$ and $\sigma_\delta^* \approx 2.5\times10^{-3}\,|\Omega|$ -- about $2\pi\times13$~kHz at $|\Omega| = 2\pi\times5$~MHz, where the time-optimal gate loses only $1.8\times10^{-5}$ to detuning noise. The correction itself is then tiny ($\sigma_\vartheta \approx 1.2\times10^{-2}$~rad).

\section{Fidelity response to laser intensity and frequency noise}\label{app:response}

\begin{figure}
\centering
    \includegraphics[width=86mm]{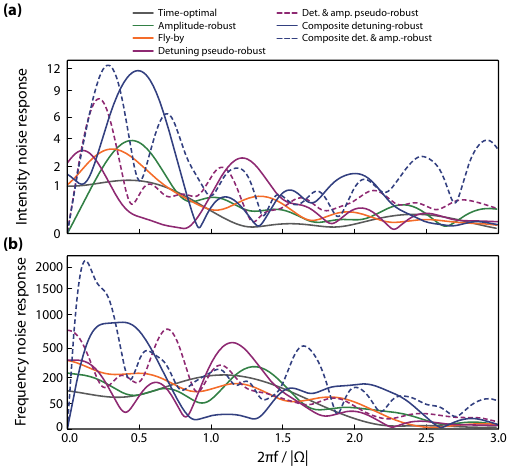}
    \caption{\label{fig:sm-response} Fidelity response functions to (a) laser intensity noise and (b) laser frequency noise for the seven protocols of Table~\ref{tab:comparison}, each at its own gate duration, as a function of the noise frequency in units of $|\Omega|/2\pi$. The value axis uses a square-root scale so that the exact zeros of the robust protocols remain visible. All curves are computed without local phase corrections; the zero-frequency values are the quasi-static sensitivities of Table~\ref{tab:comparison}, up to the amplitude-to-intensity conversion factor of $1/4$ in (a) (since $\sigma_I = 2\sigma_\epsilon$).}
\end{figure}

The quasi-static sensitivities of Table~\ref{tab:comparison} are the zero-frequency limit of a more complete characterization. Under the standard assumption that laser frequency noise and intensity noise are described by a power spectral density (PSD) whose frequency bins are uncorrelated, the gate's \emph{fidelity response} can be calculated as~\cite{tsaiBenchmarkingFidelityResponse2025}
\begin{equation}
1 - F \simeq \sum_a \int_0^\infty S_a(f)\, I_a(f)\, df,
\end{equation}
where $a$ indexes error sources (here, detuning or intensity), $S_a(f)$ is the one-sided PSD corresponding to this noise process, and $I_a(f)$ is the fidelity response function. $I_a(f)$ describes the leading-order sensitivity of the fidelity to a noise process whose PSD is a delta function at frequency $f$.

Fig.~\ref{fig:sm-response} shows $I_\epsilon(f)$ and $I_\delta(f)$ for the protocols of this work, computed from the same pulses as Table~\ref{tab:comparison} (following the convention from Ref.~\cite{tsaiBenchmarkingFidelityResponse2025} we use the laser intensity rather than the amplitude, which only modifies the function by a constant factor).

As expected, enforcing robustness to a DC error comes with trade-offs for AC errors -- i.e., when the laser frequency or intensity changes over timescales that are on the same order of magnitude as $T$. Nevertheless, we note that, using techniques like active or feedforward phase noise cancellation~\cite{Li2022ActiveCancellationServoInducedNoise,Denecker2025MeasurementFeedforwardCorrection}, AC laser noise can in principle be sufficiently reduced to be negligible for near-term experimental parameters.

\section{Principal control directions}\label{app:hessian}

\begin{figure*}
\centering
    \includegraphics[width=172mm]{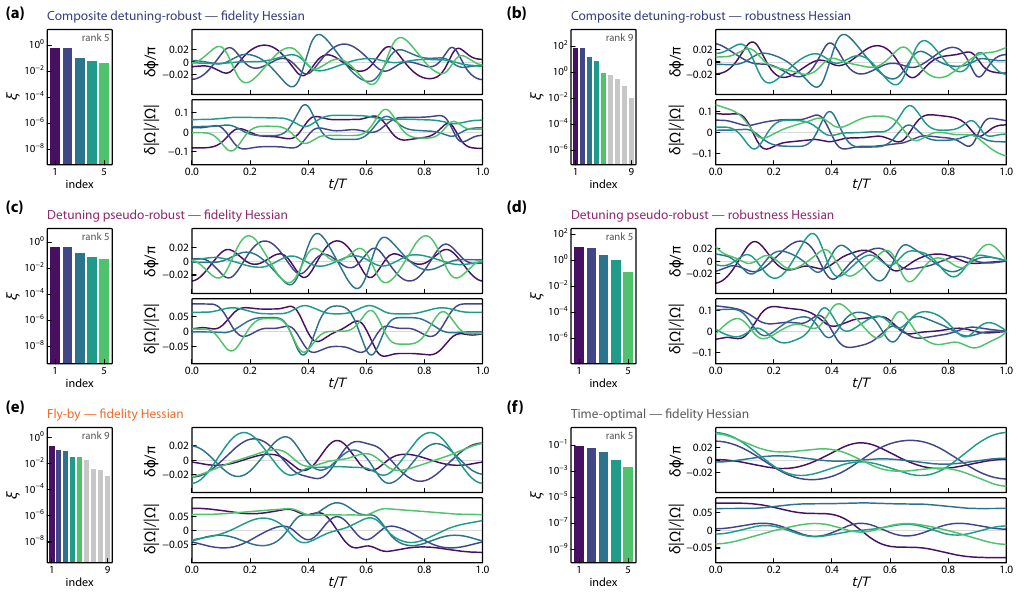}
    \caption{\label{fig:sm-hessian} Principal control directions. Each block shows, for one protocol and one Hessian (fidelity $\mathcal{H}$ or robustness $\mathcal{Q}$), the eigenvalue spectrum with its rank, and the phase ($\delta\phi$) and relative-modulus ($\delta|\Omega|/|\Omega|$) waveforms of the five leading eigenvectors (unit norm). Fidelity Hessians are shown for the composite detuning-robust, detuning pseudo-robust, fly-by and time-optimal gates; robustness Hessians for the composite and pseudo-robust gates.}
\end{figure*}

Small distortions in practical gate implementations (due to the finite modulation bandwidth, imperfect pulse shaping, etc.) can lead to significant errors, which requires \emph{in-situ} adjustments of the protocol parameters. A typical heuristic method consists in adjusting the parameters of an analytically defined modulation, but such closed form is in general not available for protocols that enforce some robustness constraint. Ref.~\cite{liu2026highfidelityneutralatomgates} shows that the high-dimensional optimization landscape (e.g., $\mathbb{R}^{2L}$ with amplitude and phase control of a length-$L$ piecewise constant pulse) can be simplified by considering the principal directions of the fidelity Hessian $\mathcal{H} = -\partial^2 F/\partial \bm{u}\,\partial \bm{u}$, such that $F(\bm{u}) \approx 1 - \frac{1}{2} \bm{u}^T \mathcal{H} \bm{u}$, where $\bm{u}$ is a small deviation from a set of control parameters that yield $F=1$. For typical Rydberg Hamiltonians, $\mathcal{H}$ has low rank (matching the intuition that conventional Rydberg protocols' errors come in a very limited number of different types), e.g. $\mathrm{rank}(\mathcal{H}) = 5$ for the symmetric infinite-blockaded Hamiltonian, and $\mathrm{rank}(\mathcal{H}) = 9$ for the fly-by infinite-blockaded Hamiltonian.

Here we extend this notion to the context of robust (or pseudo-robust) gate optimization against an error $\varepsilon \sim \mathcal{N}(0,\sigma_{\varepsilon}^2)$, where the actual average fidelity is given as
\begin{equation}
\langle F\rangle = F - \sigma_{\varepsilon}^2 S_{\varepsilon}.
\end{equation}
In this case, the infidelity under some deviation $\bm{u}$ from a set of parameters $\bm{u_0}$ that gives $F=1$ and $S_{\varepsilon} = 0$ is given by
\begin{equation}
1- \langle F(\bm{u})\rangle = \frac{1}{2} \bm{u}^T \mathcal{H} \bm{u} + \frac{1}{2}\sigma_{\varepsilon}^2 \bm{u}^T \mathcal{Q} \bm{u},
\end{equation}
where $\mathcal{Q} = \frac{\partial^2 S_{\varepsilon}}{\partial \bm{u} \partial \bm{u}}$ is the \emph{robustness Hessian}. Therefore the relevant quadratic form in that context is the total Hessian, $\mathcal{H} + \sigma_{\varepsilon}^2 \mathcal{Q}$. Finding the principal directions of this form provides the relevant modulation corrections for the total fidelity $\langle F\rangle$ that can be applied for \emph{in-situ}, error-aware, experimental optimizations.

We show in Fig.~\ref{fig:sm-hessian} the principal control directions (in amplitude and phase) for the fidelity and robustness Hessians of the composite detuning-robust gate and of the detuning pseudo-robust gate, as well as the fidelity Hessian for the fly-by and time-optimal gates. We note that $\mathrm{max}(\mathrm{rank}(\mathcal{H}), \mathrm{rank}(\mathcal{Q})) \leq\mathrm{rank}(\mathcal{H} + \sigma_{\varepsilon}^2 \mathcal{Q}) \leq \mathrm{rank}(\mathcal{H}) + \mathrm{rank}(\mathcal{Q})$, and in practice the right-hand bound is saturated (because principal directions don't align). Therefore the total Hessian has to be orthogonalized for the specific standard deviation $\sigma_{\varepsilon}$ (with total rank 14 for the composite robust gate, and 10 for the pseudo-robust gate).

\bibliographystyle{modified-ref}
\FloatBarrier
\bibliography{main}

\end{document}